# Impact of pressure and temperature on the broadband dielectric response of the HKUST-1 metal-organic framework

*Arun S. Babal,[a] Lorenzo Donà,[b] Matthew R. Ryder,[c] Kirill Titov,[a] Abhijeet K. Chaudhari,[a] Zhixin Zeng,[a] Chris S. Kelley,[d] Mark D. Frogley,[d] Gianfelice Cinque,[d] Bartolomeo Civalleri,[b] and Jin-Chong Tan[a,*]*

[a]Multifunctional Materials and Composites (MMC) Laboratory, Department of Engineering Science, University of Oxford, Parks Road, Oxford, OX1 3PJ, UK

[b]Department of Chemistry, NIS and INSTM Reference Centre, University of Turin, *via* Pietro Giuria 7, 10125 Torino, Italy

[c]Neutron Scattering Division, Oak Ridge National Laboratory, Oak Ridge, Tennessee 37831, USA

[d]Diamond Light Source, Harwell Campus, Chilton, Oxford, OX11 0DE, UK

[*]E-mail: jin-chong.tan@eng.ox.ac.uk

## Abstract

Research on the broadband dielectric response of metal-organic frameworks (MOFs) is an emergent field that could yield exciting device applications, such as smart optoelectronics, terahertz sensors, high-speed telecommunications and microelectronics. Hitherto, a detailed understanding of the physical mechanisms controlling the frequency-dependent dielectric and optical behavior of MOFs is lacking because a large number of studies have focused only on static dielectric constants. Herein we employed high-resolution spectroscopic techniques in combination with periodic *ab initio* density functional theory (DFT) calculations to establish the different polarization processes for a porous copper-based MOF, termed HKUST-1. We used alternating current measurements to determine its dielectric response between 4 Hz and 1.5 MHz where orientational polarization is predominant, while synchrotron infrared (IR) reflectance was used to probe the far-IR, mid-IR, and near-IR dielectric response across the 1.2 THz to 150 THz range (*ca*. 40 - 5000 cm$^{-1}$) where vibrational and optical polarizations are principal contributors to its dielectric permittivity. We demonstrate the role of pressure on the evolution of broadband dielectric response, where THz vibrations reveal distinct blue and red shifts of phonon modes from structural deformation of the copper paddle-wheel and the organic linker, respectively. We also investigated the effect of temperature on dielectric constants in the MHz region pertinent to microelectronics, to study temperature-dependent dielectric losses *via* dissipation in an alternating electric field. The DFT calculations offer insights into the physical mechanisms responsible for dielectric transitions observed in the experiments and enable us to explain the frequency shifts phenomenon detected under pressure. Together, the experiments and theory have enabled us to glimpse into the complex dielectric response and mechanisms underpinning a prototypical MOF subject to pressure, temperature, and vast frequencies.

---

Metal-organic frameworks (MOFs) are renowned for possessing high porosity and ordered structure along with tunable physical and chemical properties. In the past decades, the main focus of these materials was directed towards applications such as gas storage, drug delivery and chemical separations[1-4] The ability to tune the physical behavior of MOFs has opened up new avenues of research and the focus has begun to shift towards device applications, such as microelectronics, optoelectronics, luminescence and sensors.[5-9] For next-generation high-speed devices, an ultra "low-*k*" dielectric material is desirable to reduce the signal delay, power loss and electronic crosstalk with ever-shrinking device dimension and increasing number of active devices packed into an integrated circuit.[10] According to the International Technology Roadmap for Semiconductors (ITRS),[11] the conventional materials such as $SiO_2$ whose dielectric constant, static $\varepsilon'$ value or $k \sim 4$, will be replaced by a highly porous, crystalline or amorphous, chemically, and mechanically stable material in the future.[12] Due to the inherently low dielectric properties of MOFs, they could meet the timescale for the advancement of future low-*k* dielectric materials set by the ITRS.[13]

Hitherto only a handful of experiments have been reported on the dielectric properties of MOF materials. For example, Redel *et al.*[14] estimated the dielectric constant ($k$ or $\varepsilon'$) of HKUST-1 polycrystalline films in the visible wavelength range, assuming $\varepsilon' = n^2$ using the value of refractive index ($n$) measured by spectroscopic ellipsometry. Usman *et al.*[15] synthesized a thermally stable Sr-based MOF under hydrothermal conditions and measured the dielectric response of pelletized samples using impedance spectroscopy; they suggested that the low dielectric behavior is the materials intrinsic property. Zagorodniy *et al.*[16] theoretically analyzed a series of hypothetical Zn-based MOF structures using the semi-empirical Clausius-Mossotti relation and identified a number of promising candidates as ultra-low *k* dielectrics. Such theoretical work was recently extended by Ryder *et al.*[17] for a large set of carboxylate-based frameworks. Mendiratta *et al.*[18] used an experimental and theoretical approach to study the dielectric properties of a Zn-based MOF. They calculated the polarizability using the Clausius-Mossotti relation and reported that in guest-free MOF structures the main dielectric contribution is from electronic polarization.

More recently, some of us have characterized the frequency-dependent dielectric behavior of MIL-53(Al)[19] and a series of zeolitic imidazolate frameworks (ZIFs)[20] in the infrared regime. The broadband experiments were carried out using synchrotron specular reflectance spectroscopy, further corroborated by density functional theory (DFT) calculations that accurately predicted the dynamic dielectric functions observed in the experiments. For the static dielectric constant, it has been proposed that MOFs may obey a scaling rule dominated by the framework porosity and density.[17, 20] Despite these developments in the field of MOF dielectrics, there is not yet attempt to consider more rigorously the precise dielectric response and polarization mechanisms of MOFs when crossing the static ➛ kHz ➛ MHz ➛ THz frequency range. Moreover, the impact of temperature and pressure on the polarizability of MOF structures is also little understood.

In this work, we have studied the temperature- and pressure-dependent broadband dielectric behavior of an activated HKUST-1 material under vacuum conditions. The MHz measurements were carried out using an LCR meter equipped with a parallel-plate capacitor arrangement operating in the frequency range of 4 Hz to 1.5 MHz, and covering the temperature range of 20 to 100 °C. The high-frequency dielectric behavior over the far-, mid-, and near-IR regions was measured *via* synchrotron infrared specular reflectance spectroscopy covering the range of 1.2 to 150 THz (40-5000 $cm^{-1}$). Details of the experimental setups are presented in the supporting information (SI), see Figure S1. We performed density functional theory (DFT) calculations using the periodic CRYSTAL17 code[21] to gain additional insights into the effects of frequency and pressure on the broadband dielectric response of an ideal HKUST-1 structure.

The as-received Basolite C300 powder (Sigma Aldrich) also known as HKUST-1 (Figure 1a) or $Cu_3(BTC)_2$ [BTC = benzine-1,3,5-tricarboxylate] was used in this study. Water molecules can readily coordinate to the apical adsorption sites of the copper paddle-wheel,[22] hence affecting the dielectric property of the overall framework due to the strong dipole moment of water ($\varepsilon' \sim 80$ at 20 °C).[23] For this reason, in this study we have performed all dielectric measurements under vacuum conditions to activate the HKUST-1 framework, thereby eliminating the contribution of water molecules towards the total dielectric response.

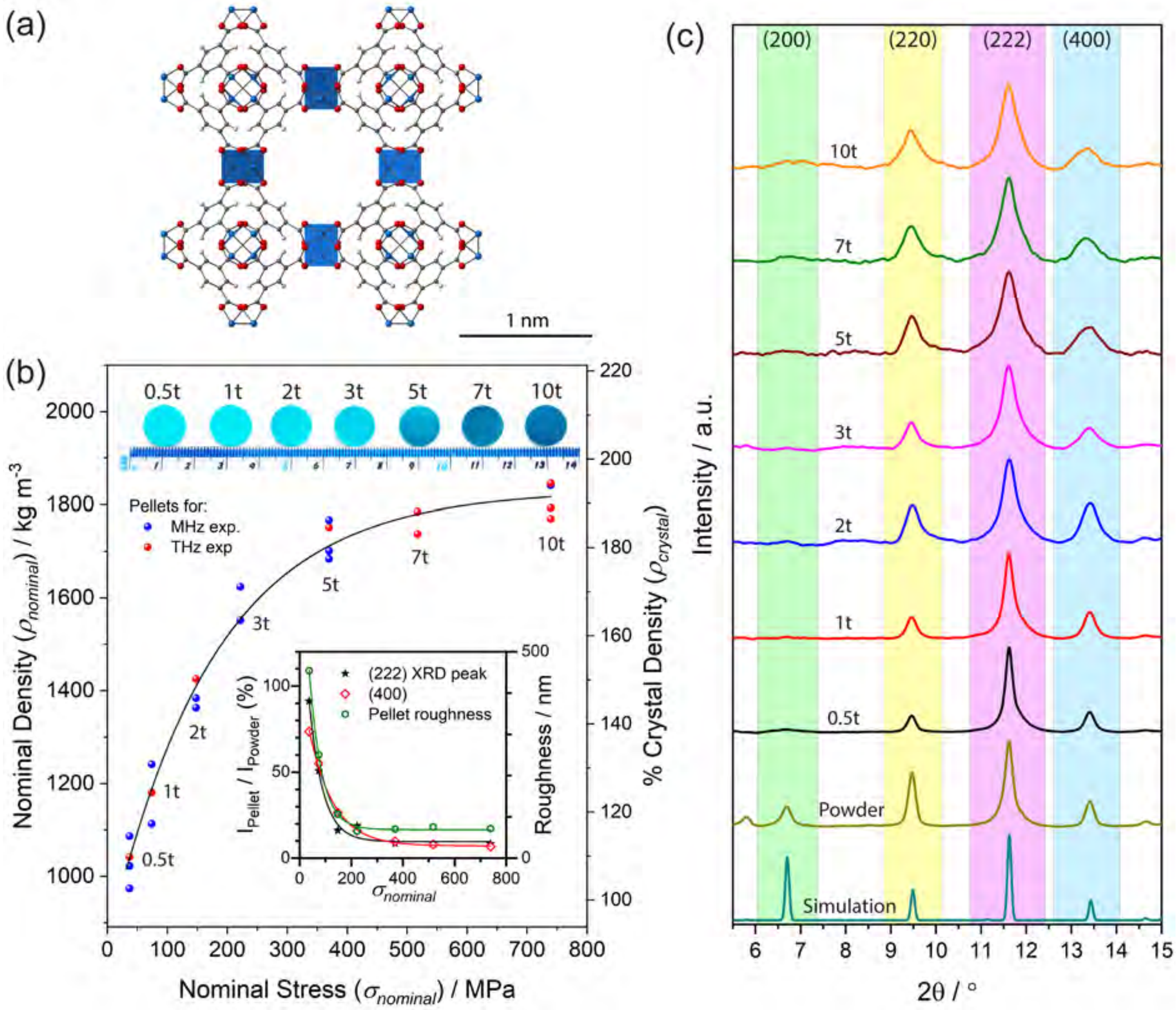

**Figure 1:** (a) Crystal structure of a unit cell of the HKUST-1 framework, shown here without adsorbed water molecules (color scheme: hydrogen in white, oxygen in red, copper in blue and carbon in grey). (b) Nominal density of the pellets used for the MHz and THz measurements plotted as a function of the nominal pellet pressure. The pellets were designated in terms of the uniaxially applied force in metric tons, e.g. 0.5t for pellet pressed with a 0.5-ton force onto a 13-mm diameter die, thus nominal stress = force / nominal area. The crystal density % was determined relative to the crystallographic density of HKUST-1. (c) Normalized XRD patterns of the HKUST-1 powder and pellets compressed under different loads. For comparison, the XRD patterns of pellets in absolute intensities are shown in Figure S3.

We have prepared pellets of HKUST-1 with a diameter of 13 mm using a manual hydraulic press, by systematically increasing the applied force from 0.5, 1, 2, 3, 5, 7 to 10 tons. Figure 1(b) shows that the prepared pellets and their nominal densities are obeying an exponential relation of the form: $\rho \propto \exp(-\sigma)$, where $\rho$ is the nominal pellet density and $\sigma$ is the nominal stress. Compared with the theoretical density of a HKUST-1 single crystal (948.9 kg $m^{-3}$),[22] the 0.5t pellet exhibits a nominal density of ~110% while the 10t pellet attained ~195% of the theoretical density. It can be seen in Figure 1(b) that the color of the pellets is also systematically shifting from turquoise towards dark blue with the rising pressure, indicating the change of its refractive index (*vide infra*).

By increasing the pelleting pressure, one may expect a reduction of free volume between the crystals in the pellet, as evidenced from roughness characterization of the polycrystalline surface, see inset of Figure 1(b) and Figure S2. Irreversible plastic deformation of the HKUST-1 structure mimicking the trend of the density vs nominal stress curve in Figure 1(b) is also expected, leading to amorphization[24] of the porous framework. X-ray diffraction (XRD) patterns in Figure 1(c) and Figure S3 show the evolution of the Bragg peaks with pelleting pressure. The HKUST-1 powder starts to amorphize under stress at 37 MPa (0.5t) as evidenced by the disappearance of the (200) and (111) peaks. The (220), (222) and (400) characteristic peaks also exhibit broadening effect with increased pelleting pressure, indicating pore collapse and framework amorphization (further supported by the full width at half maximum (FWHM) analysis in Figure S4 and Table S1). The increasing trend in FWHM values is mirroring the change of nominal density of the pellets shown in Figure 1(b).

Figure 2 shows the dielectric properties of HKUST-1 with increasing pelleting pressure and temperature at the representative frequencies of 0.01, 0.1 and 1 MHz. For an individual pellet, the real part of the dielectric constant ($\varepsilon'$) decreases with increasing frequency at a specific temperature. The MHz data revealed that the dipole moment of the framework cannot keep up with the alternating field switching at higher frequency, therefore resulting in an overall decline in the orientational polarizability of the material. For example, the $\varepsilon'$ value of the 0.5t pellet at 20 °C fell from its highest value of 2.72 at 10 kHz to 2.42 at 1 MHz. For the 10t pellet, the value of $\varepsilon'$ risen from 4.88 (20 °C) to 5.07 (80 °C) and then slightly declined to yield 4.92 at 100 °C; this effect might be linked to the negative thermal expansion of the HKUST-1 framework.[25] Higher pelleting pressure reduces the free volume between the crystals and increases amorphization of sample, both of which affect the dielectric response. For instance, for 1 MHz frequency at 20 °C, by increasing the pelleting pressure from 0.5t to 10t it was found that the value of $\varepsilon'$ can be doubled from 2.42 to 4.88. It follows that the dielectric behavior of the polycrystalline sample depends upon the pellet density, herein our data show that it scales non-linearly with pressure due to plastic deformation (see Figure 1b). This effect, however, is commonly neglected in the literature concerning the dielectric measurements of pelleted powder samples (e.g., refs. 18, 26).

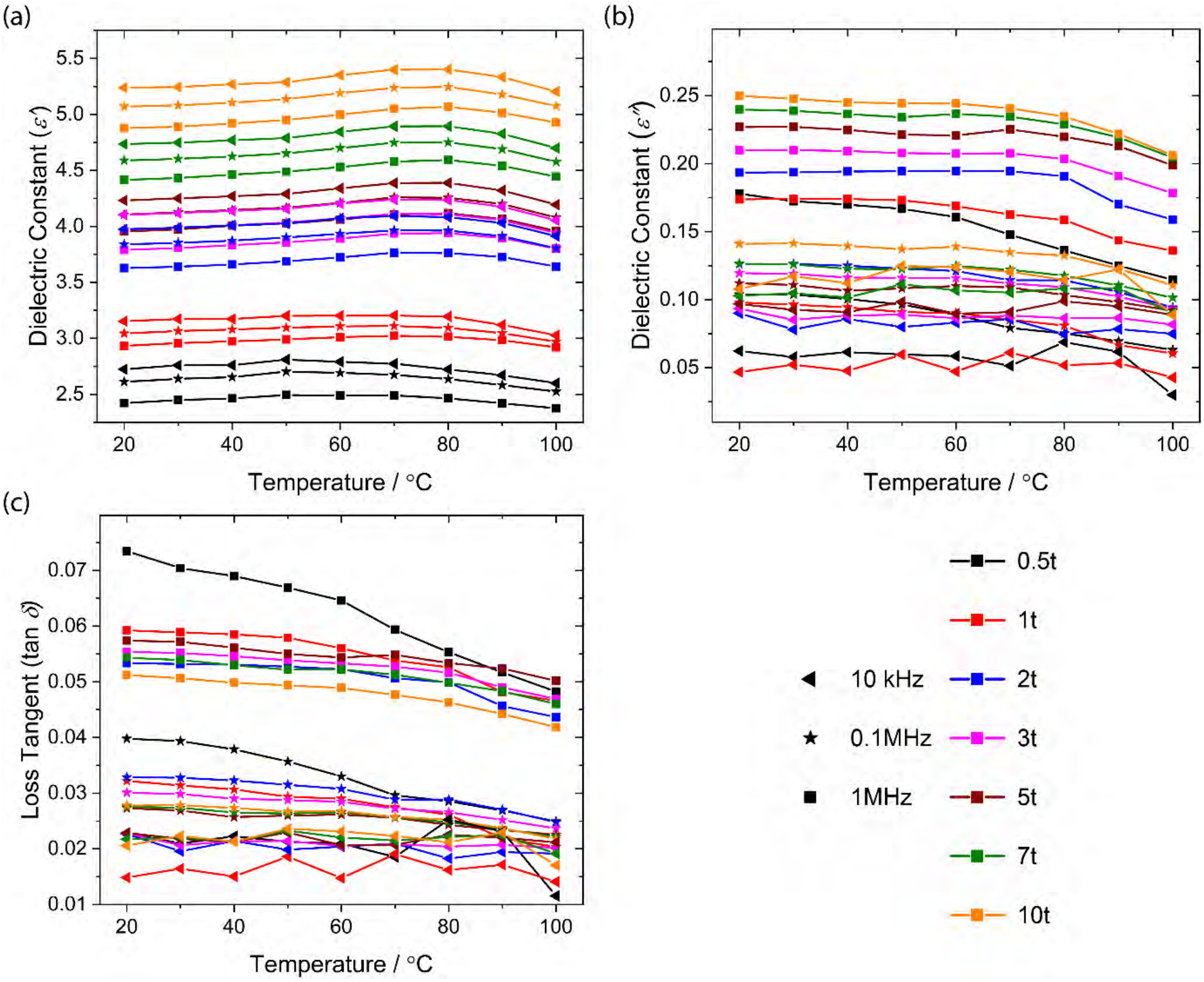

**Figure 2:** Dielectric properties of HKUST-1 as a function of temperature and pelleting pressure: (a) real part of dielectric constant $\varepsilon'$, (b) imaginary part of dielectric constant $\varepsilon''$, and (c) loss tangent, $\tan\delta = (\varepsilon''/\varepsilon')$. The marked change in the dielectric constant of the pellets from 1t to 2t can be attributed to pellet densification and framework amorphization, supported by the steep decline in relative XRD intensity and reduction of surface roughness depicted in the inset of Figure 1(b).

The imaginary part of the dielectric constant ($\varepsilon''$) is shown in Figure 2(b), which represents the energy dissipation of HKUST-1 subject to an alternating electric field. We found that all pellets have a low dielectric loss of tan $\delta$ < 0.075 (Figure 2(c)) between 20-100 °C, this value declines with rising temperature, but it increases with rising frequency. For comparison, at 1 MHz, polymers such as PVDF and PMMA exhibit a small tan $\delta$ of about 0.12[27] and 0.02,[28] respectively. It has also been reported that at 300 K a Sr-based MOF exhibits a tan $\delta$~0.02 at 0.1 MHz.[15] The complete pressure- and temperature-dependent dielectric datasets of the HKUST-1 pellets are presented in the SI, see Figures S8-S10.

Now we turn to the optical and dielectric properties of HKUST-1 in the broadband infrared (IR) frequencies, encompassing 40-5000 $cm^{-1}$. The high-resolution reflectance spectra were measured by employing synchrotron specular reflectance spectroscopy at beamline B22 in Diamond Light Source (Oxfordshire, UK); the collected reflectance spectra of HKUST-1 under different pelleting pressures are presented in Figure S11. Subsequently, the real and imaginary parts of the frequency-dependent refractive indices (Figure 3) and dielectric constants (Figure 4) were determined from the reflectance spectra by implementing Kramers-Kronig transformation,[19, 29] see full descriptions in SI section 2. To gain additional insights into the underpinning physical mechanisms, we have compared our experimental results to periodic *ab initio* DFT calculations carried out with the hybrid B3LYP-D3(ABC) functional incorporating two- and three-body dispersion forces,[30] combined with a triple-zeta quality basis set (see computational methods in SI section 10).

First, we consider the broadband optical properties of HKUST-1 as a function of frequency ($\omega$) and pelleting pressure. Figure 3 shows the real ($n$) and imaginary ($\kappa$) parts of the complex refractive index $\tilde{n}(\omega) = n(\omega) + i\kappa(\omega)$, plotted against the DFT predictions showing the general agreement between experiments and theory across the wide frequency range. The experimental spectra show that the value of $n$ varies from 1.15-1.2 for all the pellets in the NIR region (>100 THz), which is reminiscent of the value of $n$ = 1.34 calculated by DFT for the ideal HKUST-1 structure. The refractive indices increase as a function of pelleting pressure due to the increment in nominal pellet density (Figure 1(b)) and mechanically induced amorphization (Figure 1(c)), together will result in reduction of free volume ($n$ ~ 1) within a polycrystalline sample. For a particular pelleting pressure, we note that: (a) The refractive index experiences a stepwise decrease with increasing frequency, the mechanisms of which will be addressed in the sections below. (b) At each transition step, there is a discontinuity in the real part of the refractive index $n$, because of dissipative losses, giving rise to absorption peaks within the imaginary part of the refractive index $\kappa$.

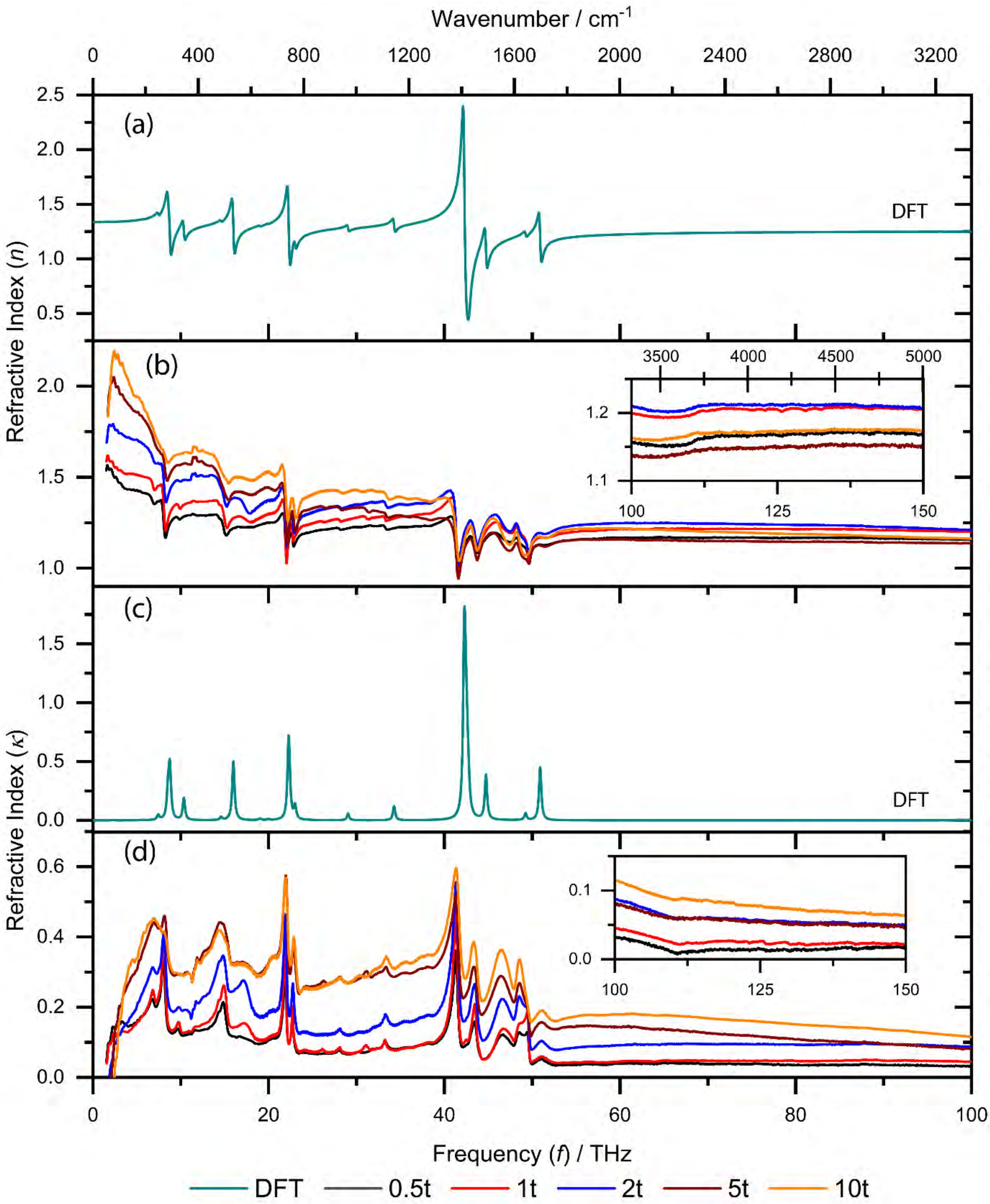

**Figure 3:** Complex refractive index of HKUST-1, where the real part of the refractive index $n$ and the imaginary part of the refractive index $\kappa$ are plotted as a function of frequency. Panels (a) and (c) are the DFT predictions of an ideal HKUST-1 structure at zero pressure. Panels (b) and (d) show the experimental results at 20 °C for the polycrystalline samples of HKUST-1 at different pelleting pressures.

Figure 4 presents the combined broadband dielectric spectra (theory vs. experiments) starting from 4 Hz up to 150 THz, to enable us to establish a complete understanding of the frequency- and pressure-dependent dielectric response of the HKUST-1 framework. To maintain consistency in the data, the broadband spectra only displays the pellet data collected at 20 °C. The complex dielectric function is $\tilde{\varepsilon}(\omega) = \varepsilon'(\omega) + i\varepsilon''(\omega)$, which comprises the real and imaginary components denoted by $\varepsilon'$ and $\varepsilon''$, respectively. It is evident from the broadband spectral data that the dielectric value decreases with increasing frequency and the static dielectric constant $\varepsilon'(\omega = 0)$ denotes the maximum value of HKUST-1.

In the near-IR region (100-150 THz), the dielectric spectra of the HKUST-1 pellets show an asymptotic behavior where $\varepsilon'$ were found to be lying in a narrow band of 1.3-1.5 (see inset of Figure 4d). This is in good agreement with the DFT values of ~1.5 beyond 100 THz. Because the dielectric mechanism in the near-IR region is attributed to femtosecond ($10^{-15}$ s) response of the electron density,[31] the dielectric constants at high frequencies are exclusively optical dielectric response that is not sensitive to the structural deformation of HKUST-1 or the pelleting pressure used.

Descending the frequency scale we arrive at the mid-IR (~10-100 THz) region, here the dielectric mechanism is controlled by polarization processes ascribed to a sub-picosecond response time of $10^{-14}$-$10^{-13}$ s, especially atomic oscillations like bending, stretching and torsional modes of molecular moieties. While in the far-IR region < 10 THz, the dielectric constants are controlled by the picosecond ($10^{-12}$ s) response of collective atomic vibrations or phonon modes, where the polarizability depends on the polarity of the chemical bonds in the framework and amplitude of the vibrations. From the theoretical DFT spectra of the ideal HKUST-1 structure, it can be seen that the resonance positions of $\varepsilon'$ (Figure 4b) and the corresponding absorption peaks of $\varepsilon''$ (Figure 4f) below 50 THz are in good agreement with the experimental spectra; also see the superimposed spectra in Figures S12 and S13. Indeed the distinctive transitions identified in the far-IR region originated from THz lattice dynamics prevalent in the HKUST-1 framework; for exemplar, the copper paddle-wheel vibrational motions and the BTC linker deformations at ~9 THz and ~16 THz, respectively.[32] Thus, in the mid- and far-IR regions, the dielectric function is strongly affected by the dynamics of the framework structure and the pellet density, as evidenced from synchrotron data shown in Figures 4(d, h) whereby the values of $\varepsilon'$ and $\varepsilon''$ systematically increased with pelleting pressure.

In the MHz frequency region, the polarization process is considerably slower with a microsecond ($10^{-6}$ s) response time. The dielectric response in this region could originate from the dipole polarization of the framework surrounding the Cu(II) paddle-wheel sites.[26, 33] We reasoned that this process is enhanced by the structural distortion of the framework, resulting in increased of net dipole moments from plastic deformation. This notion is supported by data in Figure 4(c, g), showing that the dielectric constant in the MHz range scales with pellet densification; for instance the value of $\varepsilon'$ has doubled from ~2.4 to 4.9 (at 1 MHz) when the pelleting pressure was raised from 0.5t to 10t. An additional contribution is from the orientational polarization of dipolar species, such as residual DMF solvents (see SI section 5) remained trapped in HKUST-1 after sample evacuation.

Finally, we note that the value of static dielectric constant from DFT calculations for an ideal HKUST-1 structure is 1.79, see Figure 4(a), which is resembling the previously computed values reported in literature *via* DFT methods ($\varepsilon'$ = 1.6,[17] 1.74),[26] or by using the semi-empirical Clausius-Mossotti relation ($\varepsilon'$ = 1.7).[16] In contrast, from the experimental data in Figure 4(c) we found the dielectric constant value for the 0.5t pellet to be relatively higher at $\varepsilon' \sim 3$ (at 100 Hz). This discrepancy might be attributed to two factors. (i) Framework amorphization and pellet densification will reduce the overall free volume ($\varepsilon' \sim 1$) in sample. (ii) DFT did not account for orientational (dipolar) polarization contribution caused by framework deformation, also the ideal HKUST-1 structure is free from all guest molecules. Note that the Clausius-Mossotti relation neglects both orientational and vibrational polarizations, consequently the theoretical models underestimate the dielectric permittivity.

It follows that the broadband dielectric data gathered experimentally and theoretically have enabled us to further breakdown the individual contributions, into the three main polarization mechanisms. The results for HKUST-1 are summarized in Table 1, where the total dielectric permittivity is taken as $\varepsilon'_{\text{total}} = \varepsilon'_{\text{dipole/orientational}} + \varepsilon'_{\text{atomic/vibrational}} + \varepsilon'_{\text{optical/electronic}}$.

**Table 1:** Contributions of the main polarization mechanisms to the total dielectric response of HKUST-1. The contribution of space charge is not considered in this analysis.

| Polarization mechanisms → | $\varepsilon'_{\text{dipole / orientational polarization}}$ | $\varepsilon'_{\text{atomic / vibrational polarization}}$ | | $\varepsilon'_{\text{optical / electronic polarization}}$ | $\varepsilon'_{\text{total}}$ |
|---|---|---|---|---|---|
| Pelleting pressure / MPa ↓ | MHz (4 Hz - 1.5 MHz) | Far-IR (1.5 - 20 THz) | Mid-IR (20 - 100 THz) | Near-IR (100 - 150 THz) | Entire spectrum |
| 0 (DFT)* | 0 | 0.12 | 0.18 | 1.49 | 1.79 |
| 36.96 (0.5t) | 1.32 | 0.15 | 0.14 | 1.36 | 2.97 |
| 73.92 (1t) | 1.33 | 0.2 | 0.17 | 1.48 | 3.18 |
| 147.84 (2t) | 2.05 | 0.39 | 0.26 | 1.55 | 4.25 |
| 369.60 (5t) | 2.03 | 0.77 | 0.32 | 1.30 | 4.42 |
| 739.20 (7t) | 3.17 | 0.71 | 0.44 | 1.44 | 5.76 |

**0 MPa data obtained from DFT calculations of an ideal HKUST-1 structure. We observed that the theoretical value (1.79) is reminiscent of experimental values of the 0.5t and 1t pellets upon removal of their dipolar contributions ($\varepsilon'_{\text{total}}$ - $\varepsilon'_{\text{dipole}}$), resulting in values that lie in the range of 1.65 to 1.85.*

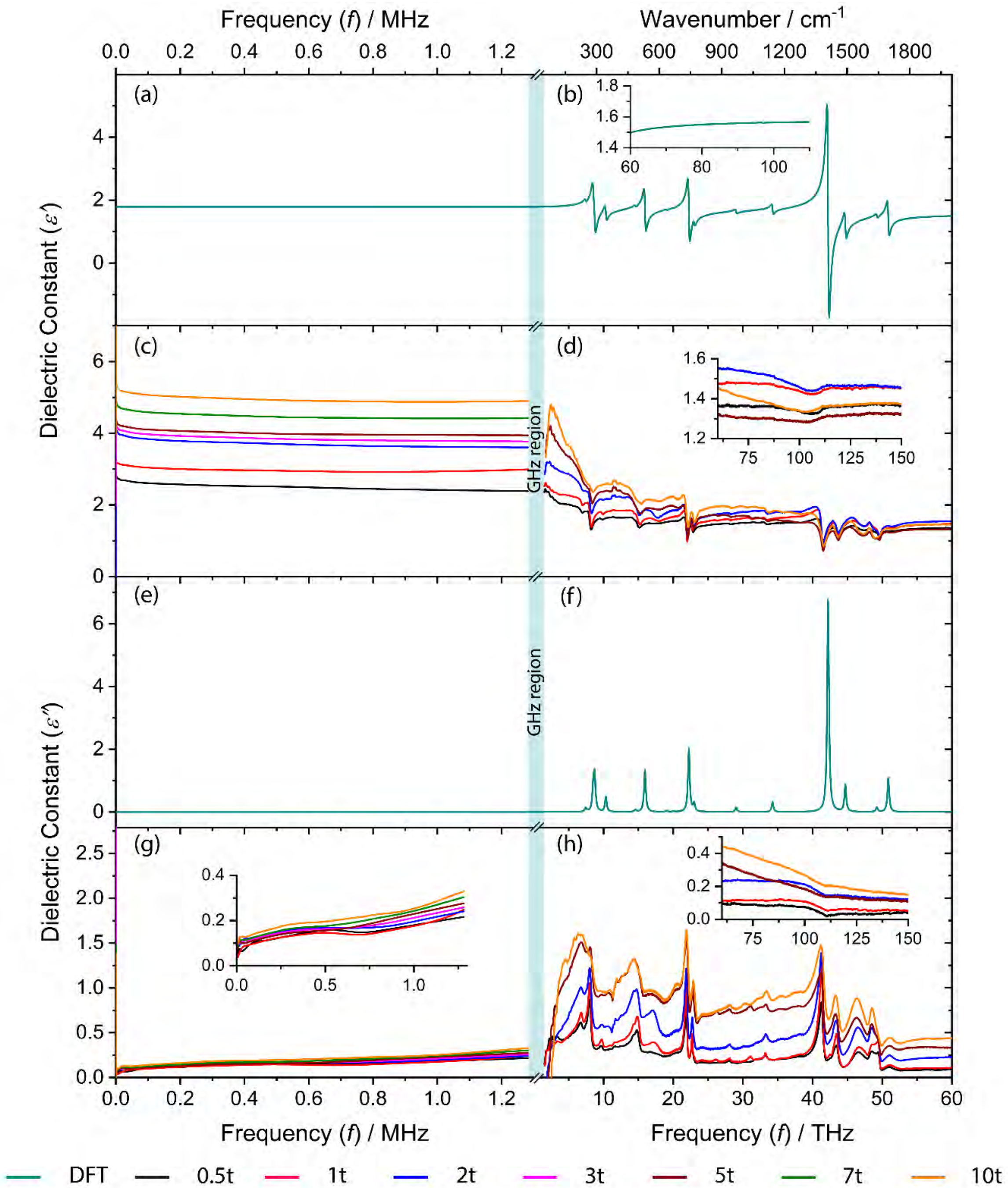

**Figure 4:** Complex dielectric function of HKUST-1 over the broadband frequency range of 4 Hz to 150 THz. (a, b) Real part of the dielectric constant $\varepsilon'$ and (e, f) imaginary part of the dielectric constant $\varepsilon''$, calculated by DFT for an ideal HKUST-1 structure at zero pressure. (c, d) Real part and (g, h) imaginary part of dielectric constants determined from experiments for pelletized HKUST-1 samples prepared under different forces. For the experimental data, the left panels show the Hz-MHz range measured by LCR parallel-plate capacitor technique, while the right panels show the far-IR, mid-IR, and near-IR regions measured by synchrotron specular reflectance spectroscopy. Note that the synchrotron measurements (i) were limited to the 0.5t, 1t, 2t, 5t and 10t pellets only, and (ii) the drop off in intensity below ~3 THz is an artefact due to frequency cutoff by the far-IR beamsplitter.

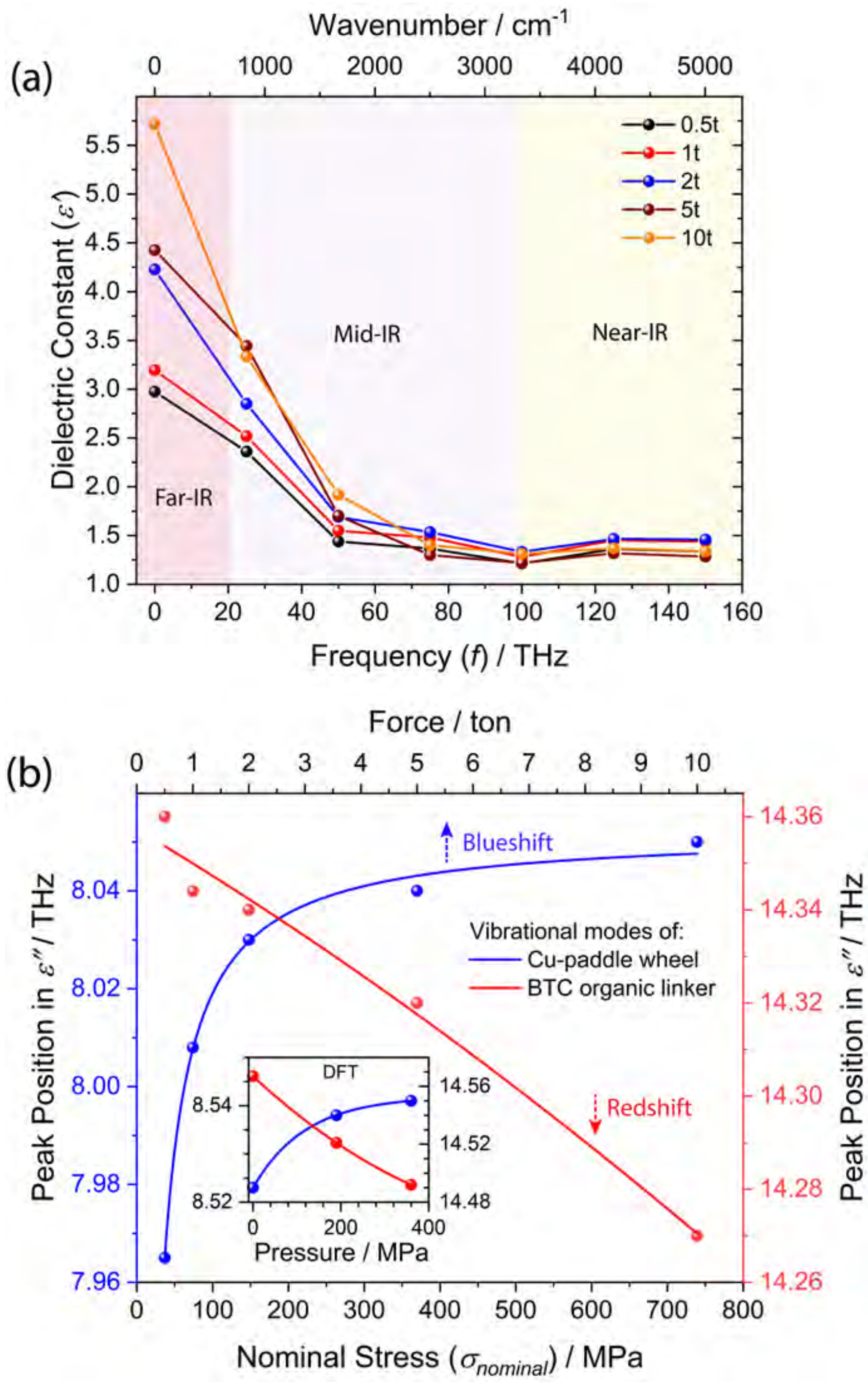

**Figure 5:** (a) Summary of the dielectric constants of HKUST-1 across the broadband frequencies comprising the far-, mid-, and near-IR regions. (b) Blue and red shifts of the THz peaks linked to the copper paddle-wheel and BTC linker vibrational modes plotted as a function of pelleting pressure; the inset shows the corresponding DFT predictions under hydrostatic pressure.

Figure 5(a) summarizes the changing trend of $\varepsilon'$ with frequency, clearly demonstrating that there exists a strong sensitivity towards the mechanically-induced structural deformation of HKUST-1 in the frequency region of below 50 THz. We note that the increase of dielectric values when progressing from the mid-IR to the far-IR region is significantly greater than those recorded through the near-IR range. Likewise, in the Hz-MHz range there is a clear dependency on mechanical stress applied during pelleting, as shown in Figure 6(a). To gain further insights we investigated the effect of mechanical deformation on the dielectric and optical properties of HKUST-1 using DFT calculations, which was achieved by imposing a hydrostatic pressure of 0, 190, and 360 MPa onto a cubic unit cell of HKUST-1, the results are shown in Figure 6(b). Whilst the cell volume declines with pressure as expected, we found that from 190 to 360 MPa the unit cell underwent a cubic to tetragonal transformation indicating mechanical instability triggered by a threshold pressure beyond ~190 MPa. In fact, we recognize that the nature of loading under nominal stress (experienced by pellets) is not identical to the hydrostatic pressure simulated by DFT, but the theoretical results shed light on the scope of employing mechanical deformation for tuning the dielectric response of a porous framework.

As the HKUST-1 framework is mechanically deformed by pellet formation, see Figure 1(b) (even at the lowest force of 0.5t), we discovered that the applied stress has resulted in THz peak shifts. Two interesting examples are shown in Figure 5(b), where the magnitude of peak shifts is plotted as a function of nominal stress. The vibrational modes at ~8 THz and ~14 THz are associated with the collective dynamics of the copper paddle-wheel (metal clusters) and the phonon modes of the BTC linkers, respectively.[32] THz vibrations of the copper paddle-wheel exhibits a blueshift with increasing stress / pressure corresponding to stiffening of the cluster deformation modes also predicted by DFT. Notably, there is a steep rise in the first ~200 MPa that can be linked to plastic deformation of the framework. On the other hand, phonon modes of the BTC linkers are softening with stress / pressure, because redshifts were detected with an increasing mechanical loading. It is quite fascinating that simulated and experimental spectra are demonstrating a similar trend (against DFT results in Figure 5 inset), except for the peak broadness of experimental data that can be attributed to polycrystalline nature of the pelletized samples. Complete analysis of the individual peak shifts in the dielectric and refractive index spectra is presented in Figures S15-S18 of the SI. Finally, the loss tangent spectra of pellets across the broadband frequencies are shown in Figure S14, demonstrating the substantially greater dissipation detected in the THz region (relative to the MHz losses) due to collective lattice dynamics and phonon vibrations of the HKUST-1 framework.

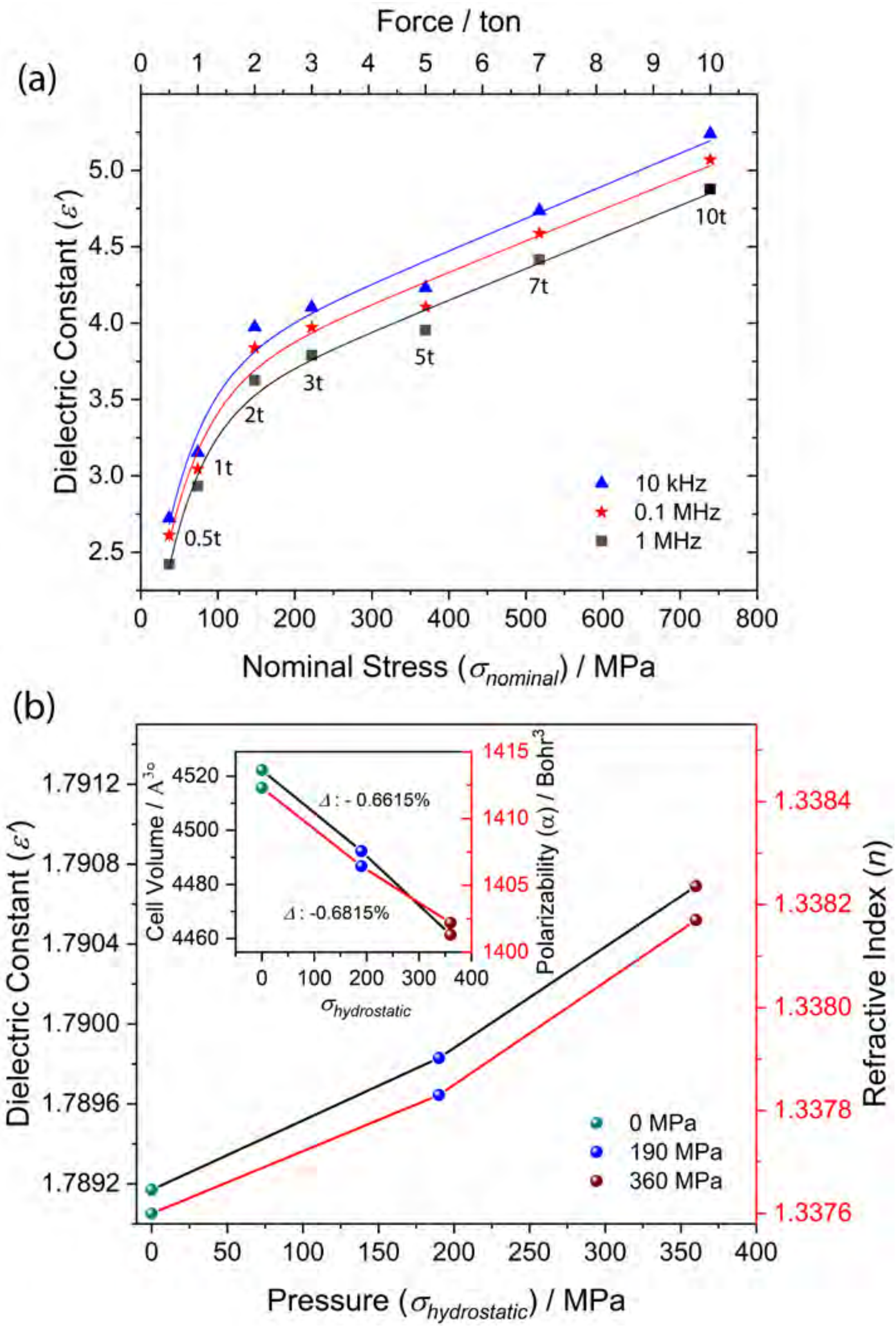

**Figure 6:** Effects of nominal stress or pressure on the dielectric constants of HKUST-1. (a) Experimental values of $\varepsilon'$ obtained from the pelletized sample measured at kHz-MHz frequencies. The curves are guides for the eye. (b) Static dielectric values computed by DFT for $\varepsilon'$ and $\varepsilon''$ at 0, 190, and 360 MPa. Inset shows the predicted change of the primitive cell volume with applied pressure, where volumetric strain is defined as $\Delta = \frac{\Delta V}{V_0} \times 100\%$. In turn, polarizability was predicted to decline with pressure, which is not unexpected because polarizability is usually proportional to the unit cell volume.

In conclusion, we reported pressure- and temperature-dependent broadband dielectric response for HKUST-1 covering the vast frequency region of 4 Hz to 150 THz. This study was made possible by employing a combination of experimental and theoretical methods. The main findings are summarized as follows:

- Dielectric constants of HKUST-1 can be modified through the application of pelleting pressure and temperature, indicating its dependency on the structural deformation, sample amorphization and densification which are associated with the free volume and framework polarizability.
- The dielectric (and optical) response of HKUST-1 in the MHz, far-IR and mid-IR regions scales very strongly with the sample microstructure, but it is independent of the framework structure within the near-IR region.
- Multiple dielectric mechanisms and polarization processes are present in HKUST-1 that can be triggered using different frequencies, namely: (i) Microsecond response in MHz region – dipole polarization of the framework and orientational polarization of guest molecules. (ii) Picosecond response in far-IR – soft modes and THz collective dynamics of flexible framework. (iii) Sub-picosecond response in mid-IR – atomic oscillations of molecular moieties of the framework. (iv) Femtosecond response from electronic polarization in near-IR – optical dielectric response not sensitive to structural deformation of the HKUST-1 framework.
- In the far- and mid-IR regions, the degree of redshift or blueshift experienced by the vibrational modes controlling the optical and dielectric response is tunable by mechanical deformation imposed by an externally applied stress / pressure. This result is confirmed by pressure-dependent DFT calculations of an ideal HKUST-1 structure.
- The dielectric values of evacuated MOF materials can be appreciably lower than conventional dielectrics (e.g., $SiO_2$, $\varepsilon' \sim 4$), they could function as a tunable dielectric for development into high-frequency applications targeting photonic sensors and THz communication devices.
- This study also led us to identifying key challenges that could hinder the use of MOFs for engineering ultra low-*k* dielectrics: (i) Complete activation of MOFs is not trivial, to eliminate trapped solvents effecting dipole polarizations at low frequencies. (ii) Porous MOF is susceptible to amorphization and structural deformation subject to thermo-mechanical stresses. (iii) MOF powders are difficult to shape for device integration.

**Supporting Information Available:** Materials characterization, TGA and FTIR analyses, pellets surface topography, reflectivity spectra, MHz and THz dielectric data, methods for DFT calculations, and detailed analyses of pressure-dependent DFT data.

**Author Contributions:** JCT conceived the project. ASB performed the MHz experiments and analyzed the dielectric data with input from JCT. LD and BC performed the DFT calculations and theoretical analysis. MRR, KT, AKC and ZXZ performed the THz synchrotron experiments with guidance from MDF and CSK, and under the supervision of GC. A.S.B. analyzed the synchrotron data with input from K.T. and J.C.T. ASB and JCT wrote the manuscript with input from all co-authors.

**Acknowledgements**:
ASB is grateful to the Engineering Science (EPSRC DTP – Samsung) Studentship that supports this DPhil research. JCT acknowledges the European Union's Horizon 2020 research and innovation programme (ERC Consolidator Grant agreement No. 771575 - PROMOFS) and the Samsung GRO Award (DFR00230) for supporting this research. M.R.R. acknowledges the U.S. Department of Energy Office of Science (Basic Energy Sciences) for research funding, and thanks the Engineering and Physical Sciences Research Council for an EPSRC Doctoral Prize Award (2017-18). We acknowledge the Diamond Light Source for the provision of beamtime SM14902 at B22 MIRIAM. We thank the Research Complex at Harwell (RCaH) for the provision of TGA and FTIR. We are grateful to Dr. Marek Jura and Dr. Gavin Stenning at R53 Materials Characterisation Laboratory in ISIS RAL for access to the XRD facilities.

# Impact of pressure and temperature on the broadband dielectric response of the HKUST-1 metal-organic framework

*Arun S. Babal,[a] Lorenzo Donà,[b] Matthew R. Ryder,[c] Kirill Titov,[a] Abhijeet K. Chaudhari,[a] Zhixin Zeng,[a] Chris S. Kelley,[d] Mark D. Frogley,[d] Gianfelice Cinque,[d] Bartolomeo Civalleri,[b] and Jin-Chong Tan[a,*]*

[a]Multifunctional Materials and Composites (MMC) Laboratory, Department of Engineering Science, University of Oxford, Parks Road, Oxford, OX1 3PJ, United Kingdom

[b]Department of Chemistry, NIS and INSTM Reference Centre, University of Turin, *via* Pietro Giuria 7, 10125 Torino, Italy

[c]Neutron Scattering Division, Oak Ridge National Laboratory, Oak Ridge, Tennessee 37831, United States

[d]Diamond Light Source, Harwell Campus, Chilton, Oxford, OX11 0DE, United Kingdom

*E-mail: jin-chong.tan@eng.ox.ac.uk

# Table of contents

# 1. Material and characterization

## 1.1 Material

The HKUST-1 (copper benzene-1,3,5-tricarboxylate) MOF material also known as Basolite C300 was purchased from Sigma Aldrich and used without further purification.

## 1.2 Pellet preparation

The HKUST-1 pellets with a diameter of 13-mm were prepared under different mechanical loads (0.5, 1, 2, 3, 5, 7 and 10 tons force) using a hydraulic press (Specac 15 tons capacity). The pellet mass was kept constant at 150 mg. The pellets were designated as *N*t, where *N* is the applied mechanical load during pelleting. The calculated nominal density was the ratio of individual pellet mass to the pellet volume.

## 1.3 Optical microscopy and surface profilometry

Alicona profilometer was used to measure the surface texture such as roughness of the MOF pellet. The surface topography was characterized by infinite focus microscopy technique (Alicona Infinite Focus 3D profilometer) using the 20× optics on the profilometer.

## 1.4 Fourier-transform infrared (FTIR) spectroscopy

The FTIR spectra of the pellets were characterized in the mid-IR region at 0.5 $cm^{-1}$ spectral resolution by employing the Nicolet-iS10 FTIR spectrometer equipped with an attenuated total reflection (ATR) accessory.

## 1.5 Thermogravimetric analyses (TGA)

The thermal stability of the HKUST-1 specimens was measured using the TGA-Q50 (TA Instruments) equipped with an induction heater (max temperature 1000 °C) and platinum sample holder under an $N_2$ inert atmosphere. The samples were heated at a rate of 10 °C/min from 40 to 500 °C.

## 1.6 X-ray diffraction (XRD)

The XRD patterns of HKUST-1 powder and pellets were measured using the Rigaku Miniflex bench-top X-ray diffractometer. The data were collected from the Bragg diffraction angle of $2\theta = 2°$ to 30°, at a scan rate of 1°/min with a step size of 0.05°.

# 2. Experimental setup and analysis of dielectric and reflectivity data

### 2.1 Using the LCR meter to measure Hz-MHz range

The dielectric response of different pressure pellets was measured using the Hioki-IM3536 LCR meter in the frequency range of 4 Hz to 1.5 MHz, see Figure S1(a). The measurements were based on the principle of a parallel plate capacitor. The two opposing surfaces of each pellet sample coming into contact with the electrodes were coated with a thin layer of silver conducting paint (RS). The parallel plate setup was placed in a vacuum oven with a 50-L chamber, and calibrated (at both open and closed circuit) beforehand in order to eliminate any parasitic impedance and admittance. Before pellet measurements, each sample was first evacuated in the vacuum ($10^{-3}$ bar) for 16 hours and thereafter temperature dependent measurements were carried out with a step increase of 10 °C under the same vacuum environment. The real ($\varepsilon'$) and imaginary ($\varepsilon''$) parts of the dielectric constant as a function of frequency ($\omega$) was calculated from the following equations:

$$\varepsilon'(\omega) = \frac{C(\omega)d}{\varepsilon_0 A}$$

$$\varepsilon''(\omega) = \varepsilon'(\omega)\tan\delta$$

where $C$ is the capacitance, $d$ is the distance between the pair of parallel plate electrodes, $A$ is area of the electrode, $\varepsilon_0$ is vacuum permittivity and $\tan\delta$ is the loss tangent.

### 2.2 Using synchrotron infrared (IR) specular reflectance spectroscopy to measure the far- (THz), mid- and near-IR frequencies

The IR reflectivity spectra of pellets prepared using compression varied from 0.5 to 10 tons were carried out at Beamline B22 MIRIAM in the Diamond Light Source (Harwell, UK), see Figure S1(b). We employed the Bruker Vertex 80V FTIR interferometer equipped with the Pike Technologies VeeMAX II variable angle specular reflectance accessory to measure the specular reflectance spectra under vacuum conditions at room temperature (21 °C), in the frequency range of 1.5 THz to 150 THz. The specular reflection spectra were collected at an angle of 30° from the normal axis of the pellet surface at 2 $cm^{-1}$ resolution and 512 scans per spectral scan. The 6-µm thick Mylar broadband multi-layer coated beam splitter was used to perform the measurement in far-IR, whereas KBr beam splitter was used for the mid-IR region. The background spectra were collected before each far and mid-IR measurements.

The Kramers-Kronig Transform (KKT) [1] is based on the causality principle that shows the dependency of the real and imaginary part of complex quantities i.e. dielectric constant ($\tilde{\varepsilon}(\omega) = \varepsilon'(\omega) + i\varepsilon''(\omega)$), refractive index ($\tilde{n}(\omega) = n(\omega) + i\kappa(\omega)$) and logarithm of the amplitude-reflectivity on each other. The built-in KKT routine in the Bruker OPUS software only takes piecewise information from the reflectance spectra resulted into the negative values for the imaginary part of refractive index and dielectric constant. The mid- and far-IR reflectivity spectra were smoothly joined together using the MATLAB code to get a continuous reflectance spectra as described in ref. [2].

The phase change ($\phi$) at an arbitrary wavenumber ($\omega_a$) was calculated using the Kramers-Kronig relation:

$$\phi(\omega_a) = \frac{2\omega_a}{\pi}\int_0^\infty \frac{\log\left(\sqrt{R(\omega)}\right)}{\omega^2 - \omega_a^2} d\omega$$

The real $n(\omega)$ and imaginary $\kappa(\omega)$ parts of the complex refractive index are given by [3]:

$$n(\omega) = \frac{1 - R(\omega)}{1 + R(\omega) - 2\sqrt{R}\cos(\phi(\omega))}$$

$$\kappa(\omega) = \frac{-2\sqrt{R}\sin(\phi(\omega))}{1 + R(\omega) - 2\sqrt{R}\cos(\phi(\omega))}$$

The relation between complex dielectric constant $\tilde{\varepsilon}(\omega)$ and refractive index $\tilde{n}(\omega)$ can thus be obtained as:

$$\tilde{\varepsilon}(\omega) = \tilde{n}(\omega)^2$$

$$\varepsilon'(\omega) = n(\omega)^2 - \kappa(\omega)^2$$

$$\varepsilon''(\omega) = 2n(\omega)\kappa(\omega)$$

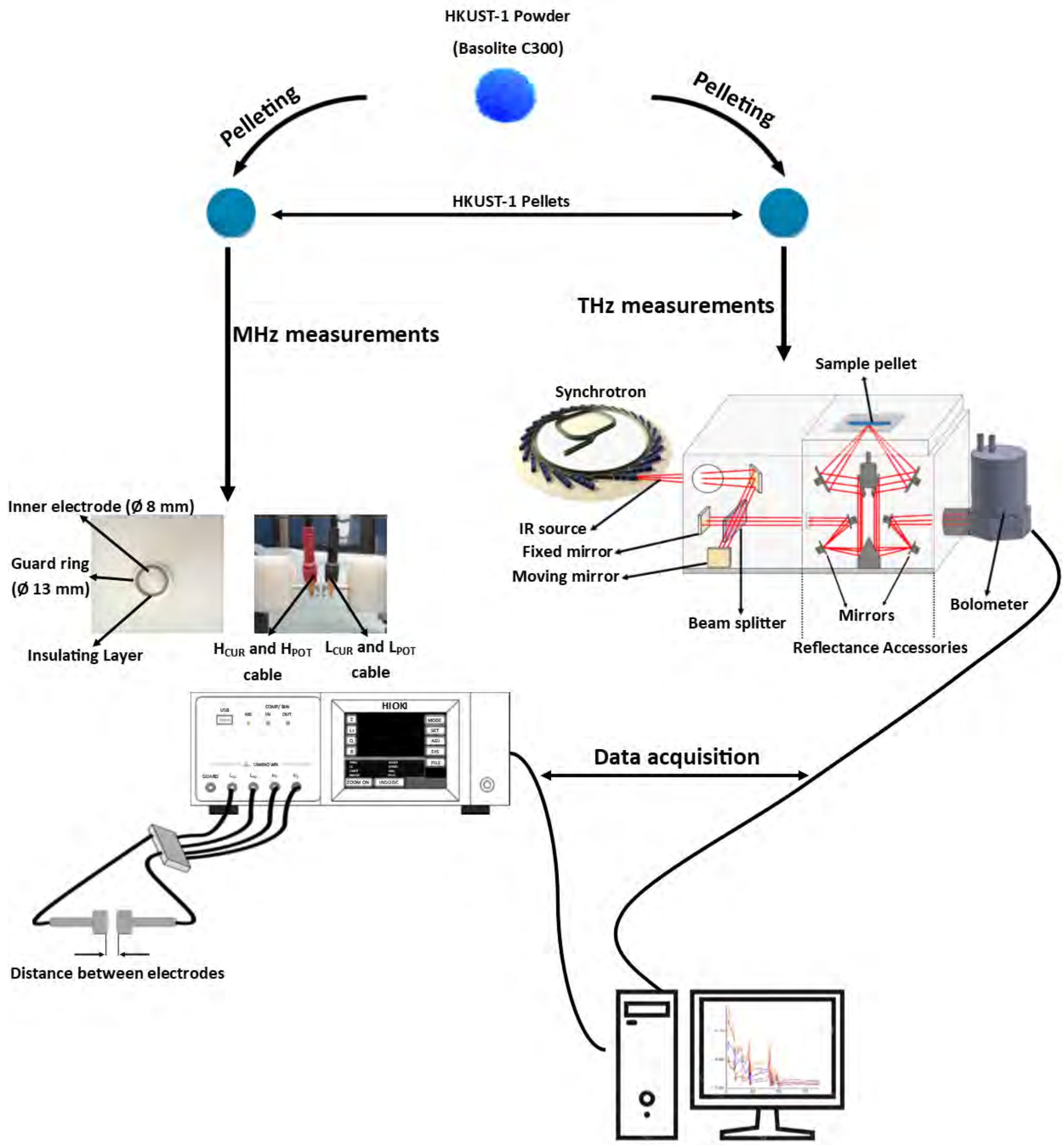

Figure S1: Experimental setups used to conduct the broadband dielectric measurements on the HKUST-1 MOF pellets.

## 3. Pellets topographic characterization

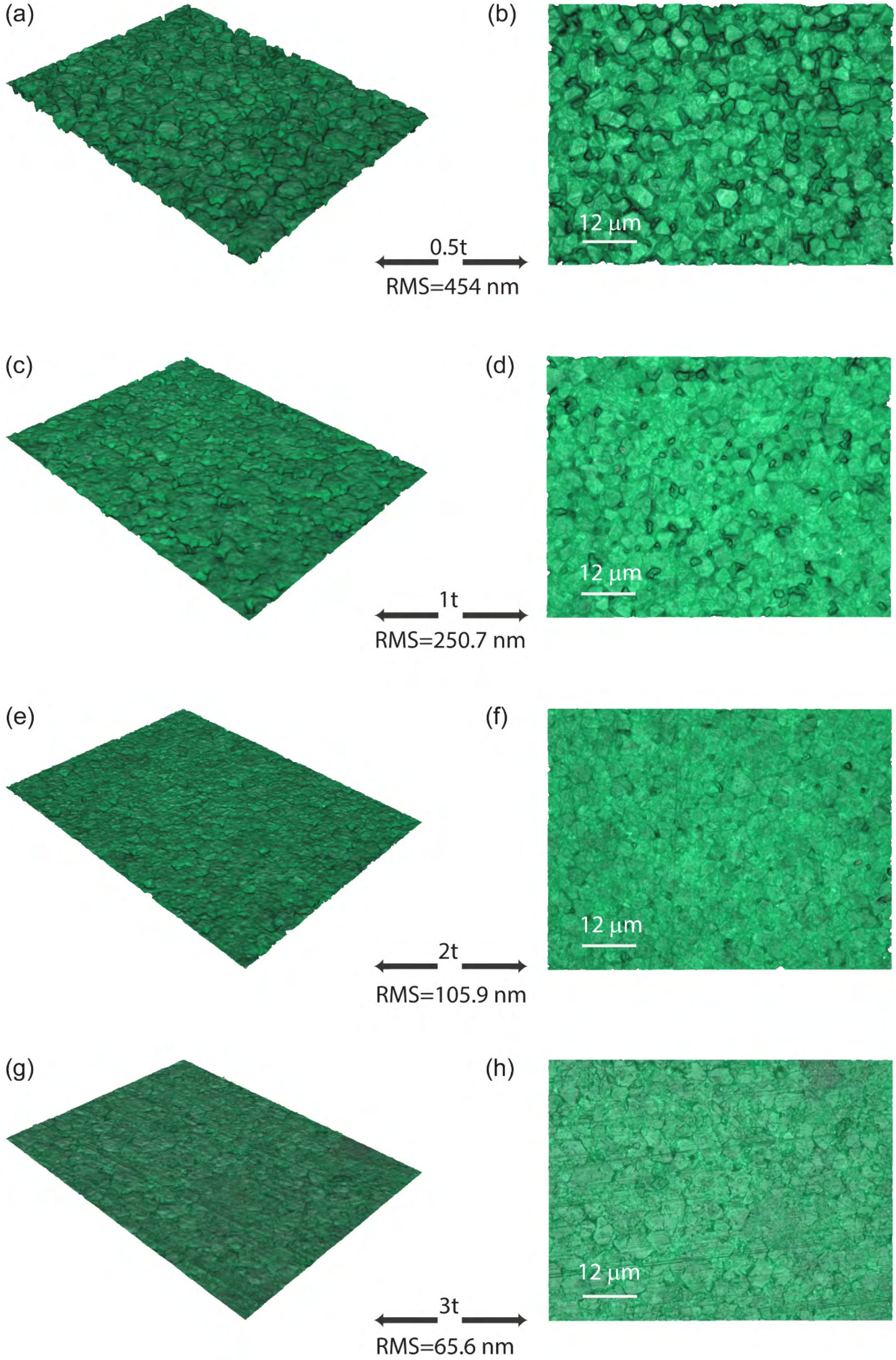

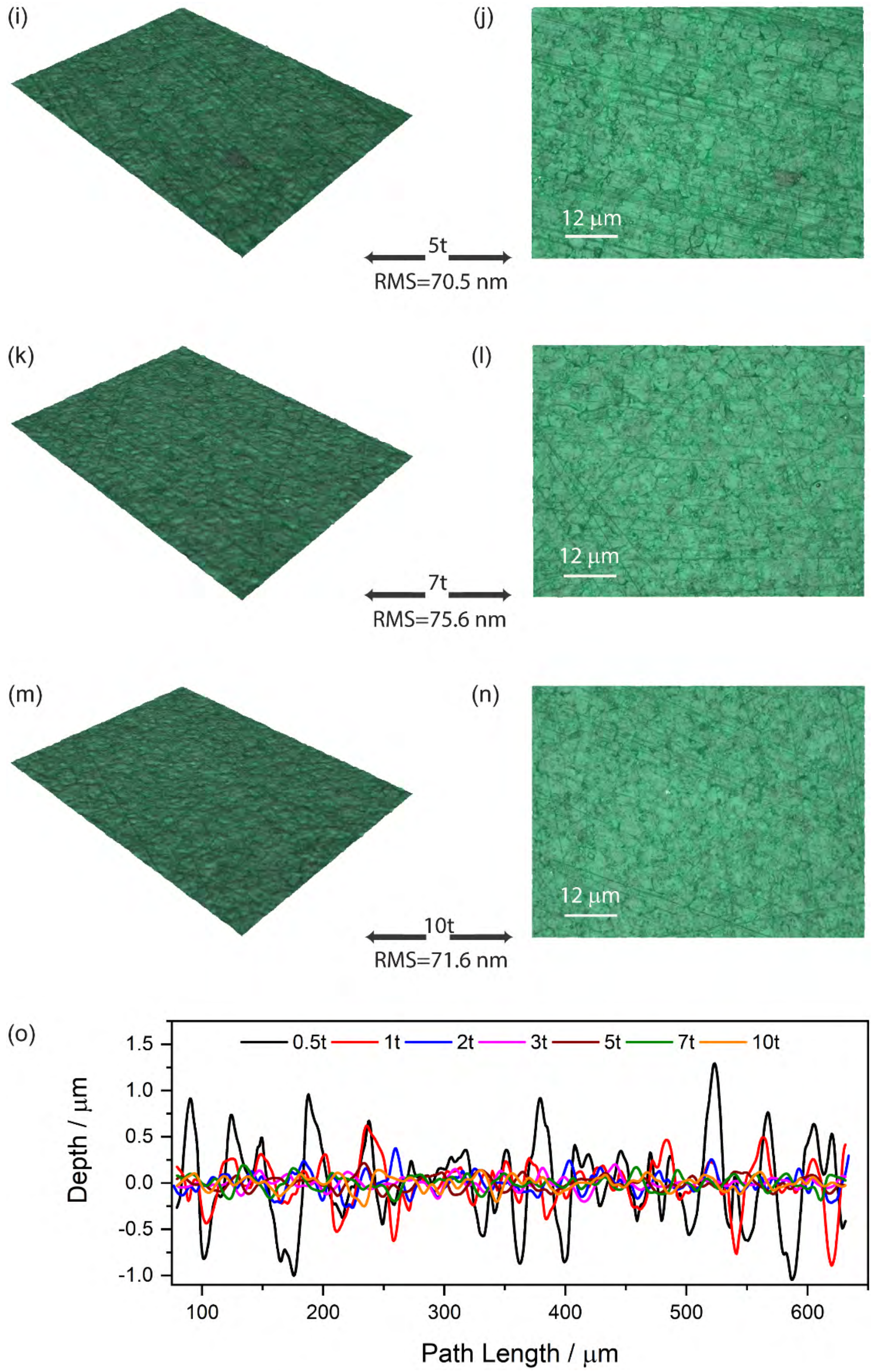

Figure S2: Pellet surface topography characterized using the Alicona Infinite Focus profilometer at 20× optical magnification: (a)-(b) 0.5 ton, (c)-(d) 1 ton, (e)-(f) 2 ton, (g)-(h) 3 ton, (i)-(j) 5 ton, (k)-(l) 7 ton, (m)-(n) 10 ton and (o) Pellet roughness profile (the width of profile line is 80 μm), respectively. RMS represents the root mean squared roughness value of the surface profile.

## 4. Effect of nominal pressure on FWHM of the XRD reflections

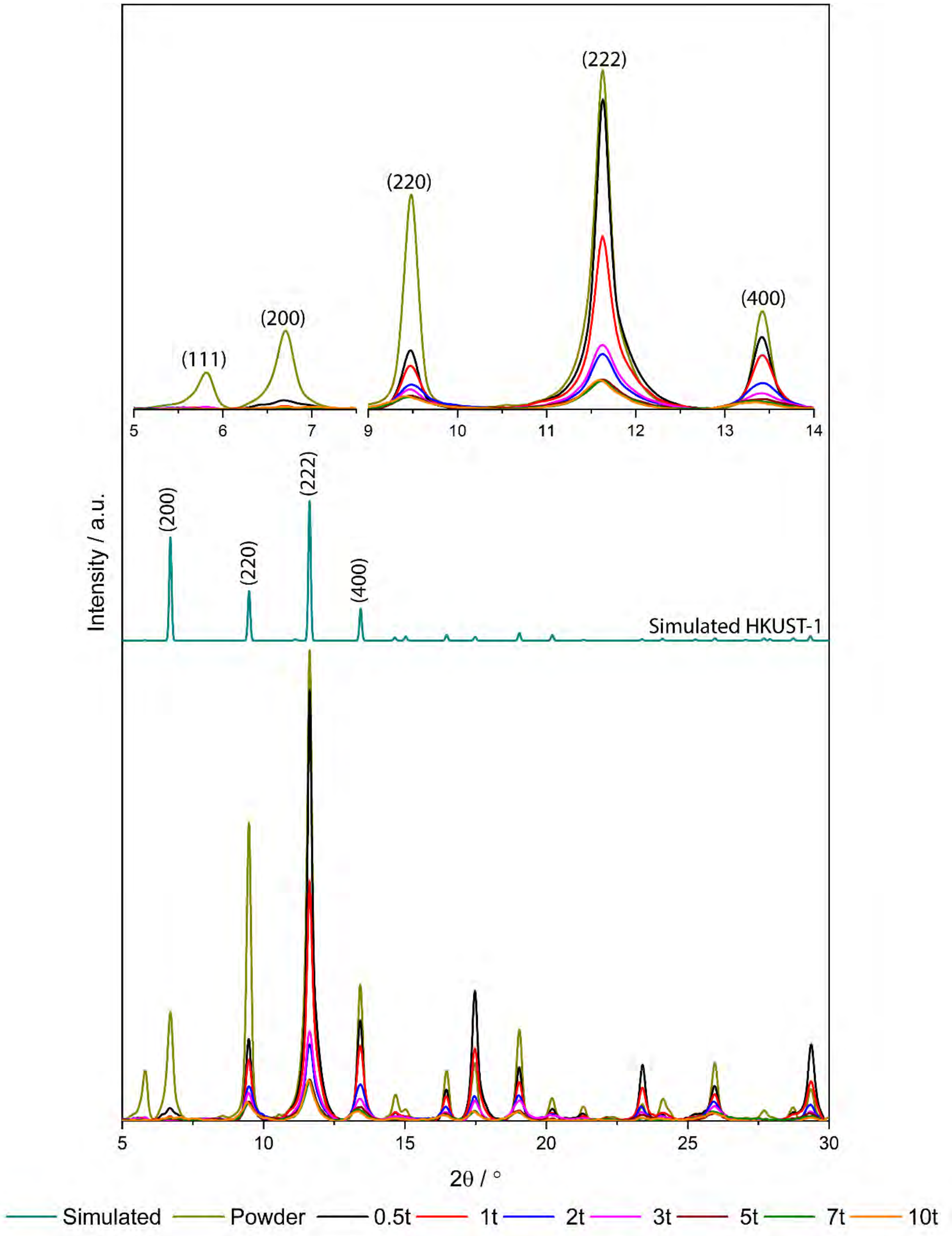

Figure S3: Non-normalized XRD patterns of HKUST-1 samples in absolute intensities. Inset: magnified view of the main characteristic peaks.

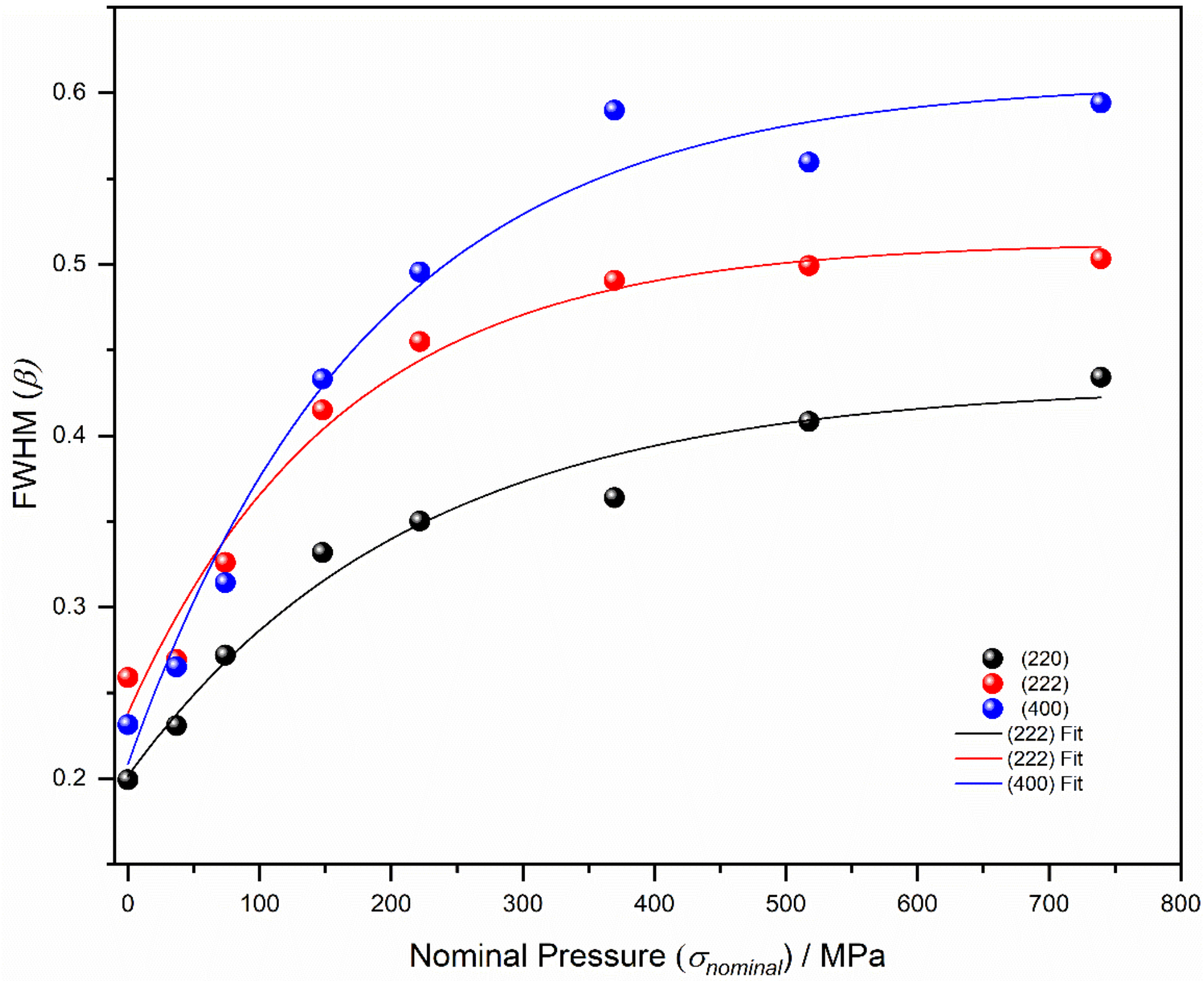

Figure S4: FWHM plot of HKUST-1 MOFs XRD peaks as a function of applied pressure. In this plot effect of pressure on various characteristic planes i.e. (220), (222) and (400) was investigated using the full width at half maximum (FWHM). The curves are following a similar pattern as the nominal density plot indicates that high compression loading is resulting in MOF framework amorphization.

Table S1: FWHM values of XRD planes for different pressure pellets.

| Designation | Pressure (MPa) | Characteristics planes of HKUST-1 | | |
|---|---|---|---|---|
| | | (220) | (222) | (400) |
| Powder | 0 | 0.20 | 0.26 | 0.23 |
| 0.5t | 36.96 | 0.23 | 0.27 | 0.27 |
| 1t | 73.92 | 0.27 | 0.33 | 0.31 |
| 2t | 147.84 | 0.33 | 0.42 | 0.43 |
| 3t | 221.76 | 0.35 | 0.45 | 0.50 |
| 5t | 369.60 | 0.36 | 0.49 | 0.59 |
| 7t | 517.44 | 0.41 | 0.50 | 0.56 |
| 10t | 739.20 | 0.43 | 0.50 | 0.59 |

## 5. TGA Data and Analysis

TGA was carried out to measure the thermal stability of HKUST-1 pellets (see Figure S4). The total weight loss of pellets decreases as a function of pressure. The initial weight loss up to 150 °C is assigned to the $H_2O$ desorption resulted from the hydrophilic nature of HKUST-1. The presence of solvent in the sample was confirmed by another weight loss in the range of 200-300 °C, which is assigned to the removal of entrapped DMF from solvothermal synthesis. Finally, the sharp weight loss at ~350 °C in all the samples is corresponding to the structural decomposition of the HKUST-1 framework.

Table S2: Number of DMF molecules (*n*) trapped in Basolite C300 (HKUST-1) estimated from the TGA data, where $[Cu_3(BTC)_2] \cdot nC_3H_7NO$.

| Pellet sample | *n* |
|---|---|
| 0.5t | - |
| 1t | - |
| 2t | 0.24 |
| 3t | 0.24 |
| 5t | 0.24 |
| 7t | 0.19 |
| 10t | 0.19 |

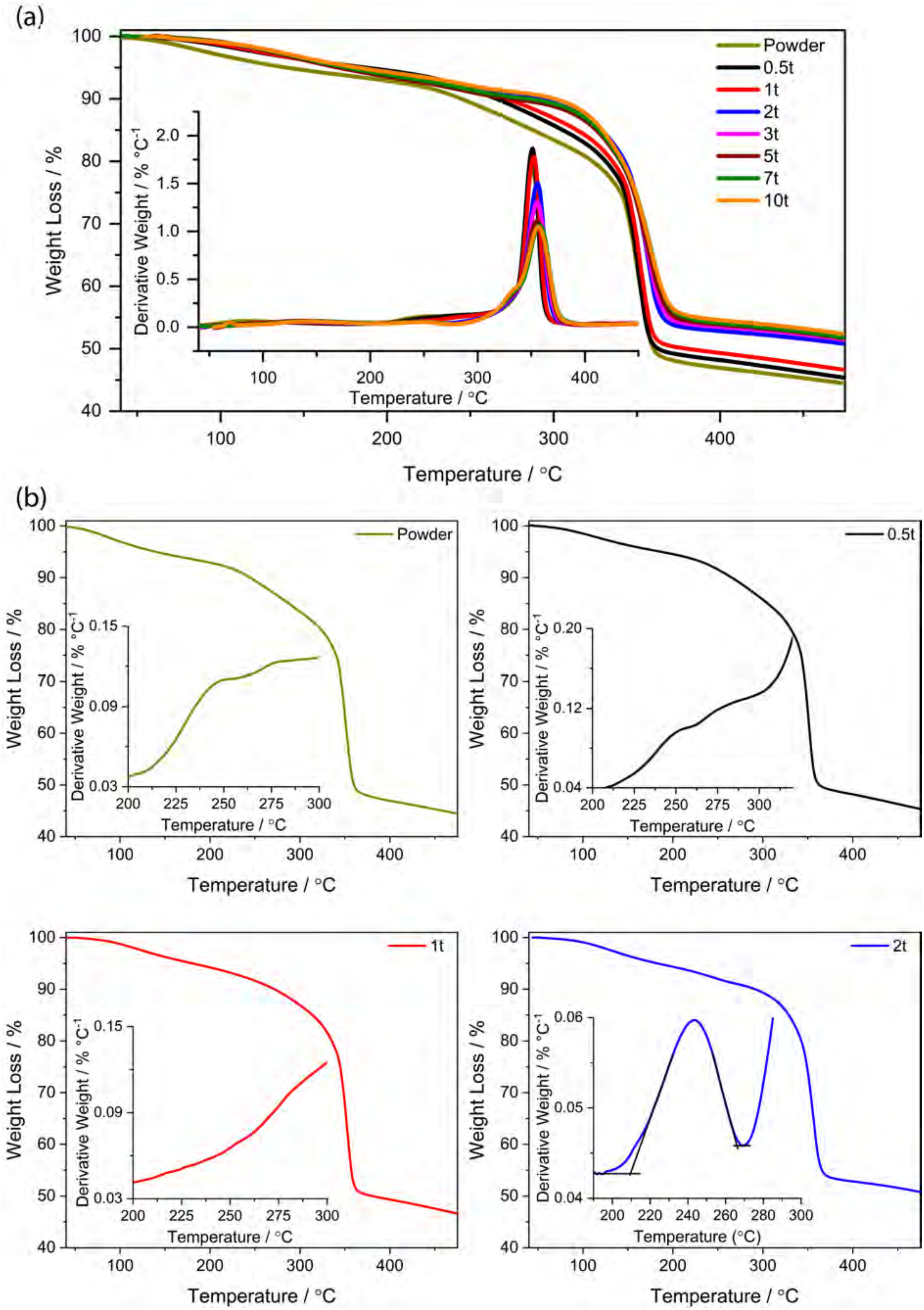
(a)
Powder
0.5t
1t
2t
3t
5t
7t
10t
Weight Loss / %
Derivative Weight / % °C-1
Temperature / °C
(b)
Powder
0.5t
1t
2t
Temperature (°C)

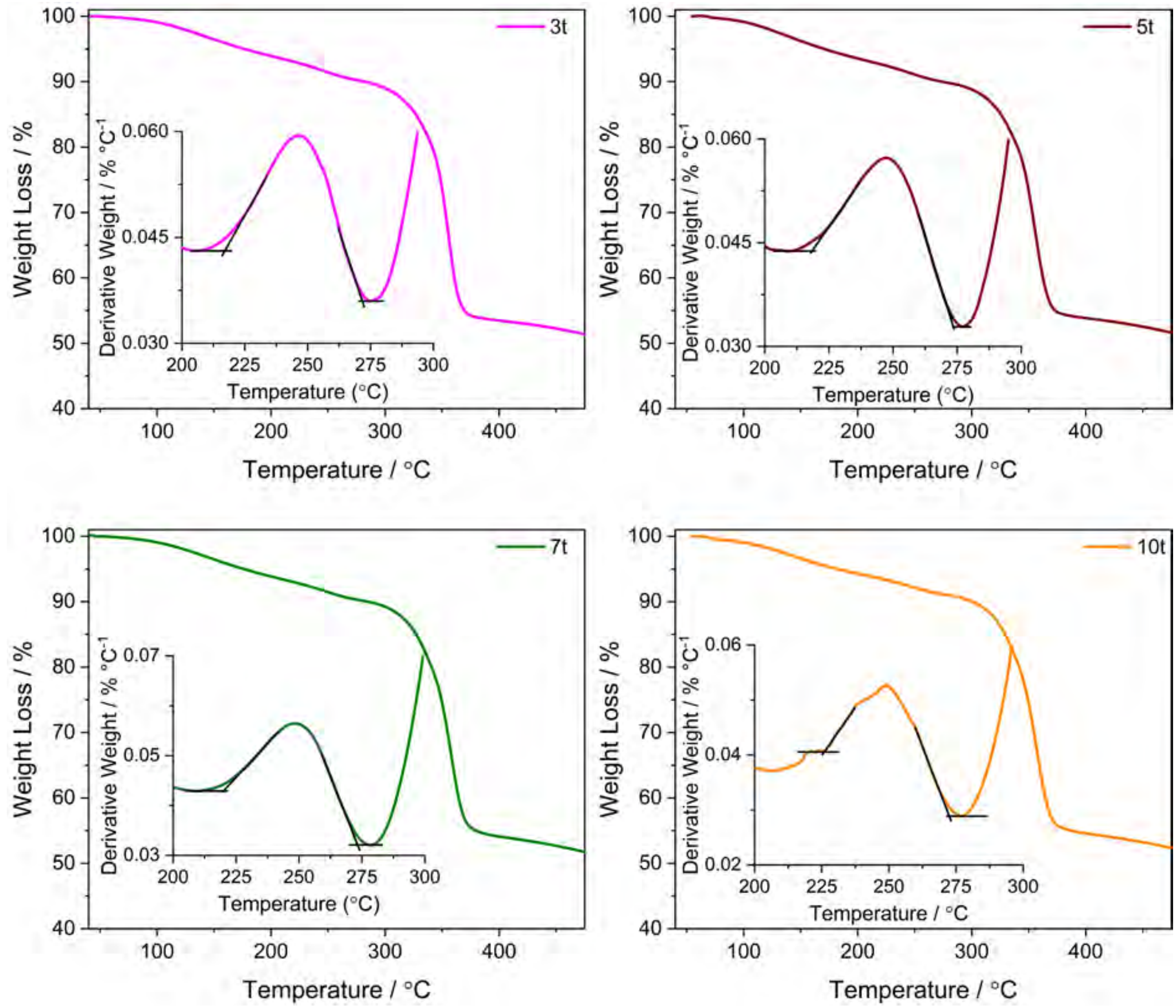

Figure S5: (a) TGA of HKUST-1 samples (Basolite C300) as a function of temperature. Inset shows the derivative weight change as a function of temperature. The hump in the derivative plots (b) of pellets suggest the presence of some DMF molecules from solvothermal synthesis remain trapped in the framework. With the effect of pelleting pressure the weight loss associated with DMF and organic linker becomes more distinctive, which was found overlapped in the pristine powder of HKUST-1.

## 6. FTIR Data and Analysis

The presence of residual DMF solvent in the MOF framework was further confirmed by the ATR-FTIR spectroscopy shown in Figures S6 and S7. Peaks at 1373 and 1548 $cm^{-1}$ correspond to the C=C stretching whereas 1450 and 1648 $cm^{-1}$ are associated with the symmetric and asymmetric COO stretching, respectively. It can be seen that these peaks broaden with pelleting pressure as a result of amorphization [4]. The peaks at 1260 and 2925 $cm^{-1}$ are assigned to the asymmetric and symmetric C-N stretching, which are related to the trace amount of DMF solvent trapped inside the framework.

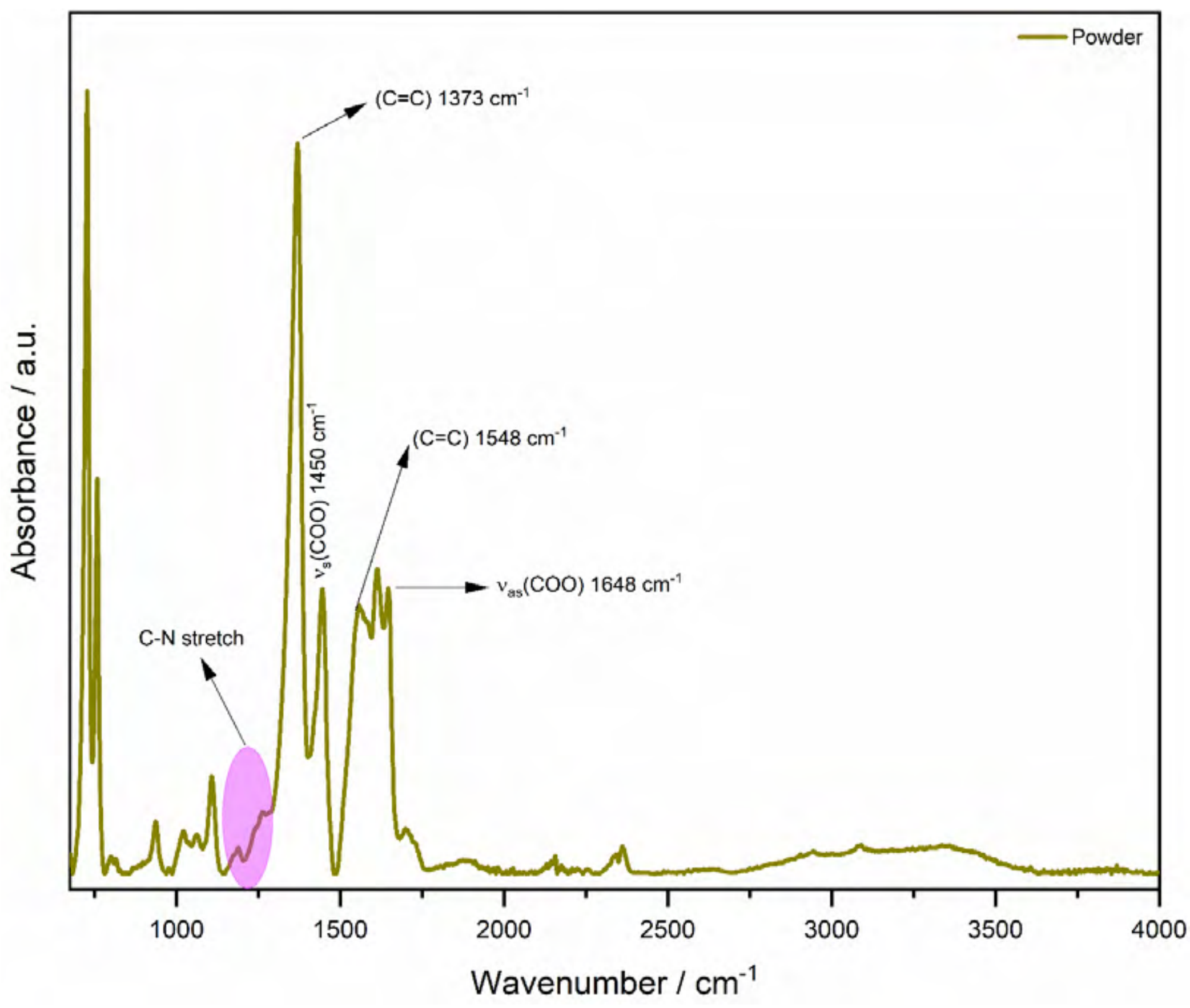

Figure S6: FTIR spectra of HKUST-1 powder. ($\nu_s$=symmetric stretching and $\nu_{as}$=asymmetric stretching).

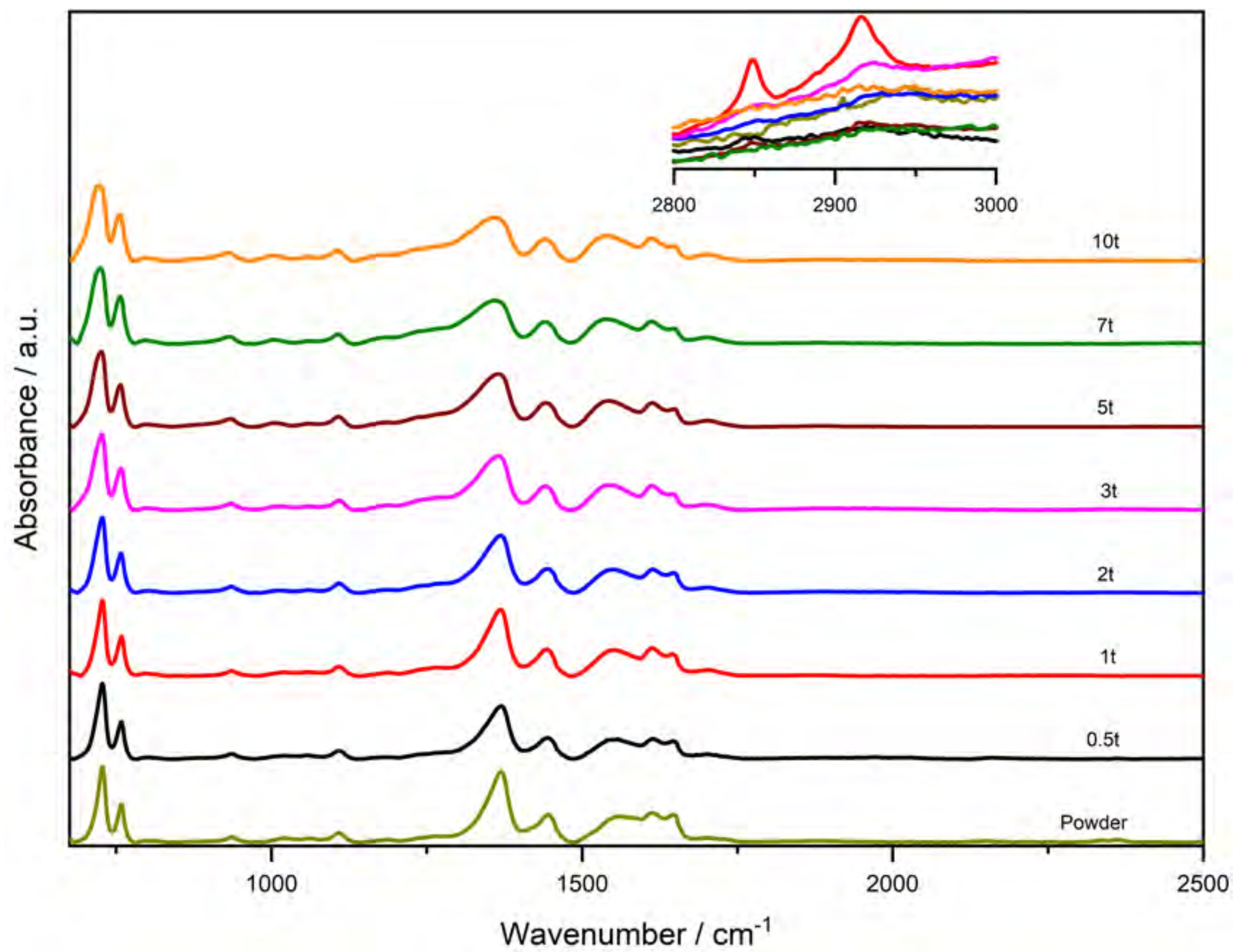

Figure S7: FTIR spectra of HKUST-1 pressure pellets. The inset demonstrates the spectra at higher wavenumber. The frameworks characteristic peaks show broadening with the applying pressure.

## 7. Real part of dielectric constant in MHz region - Individual pellet dielectrics

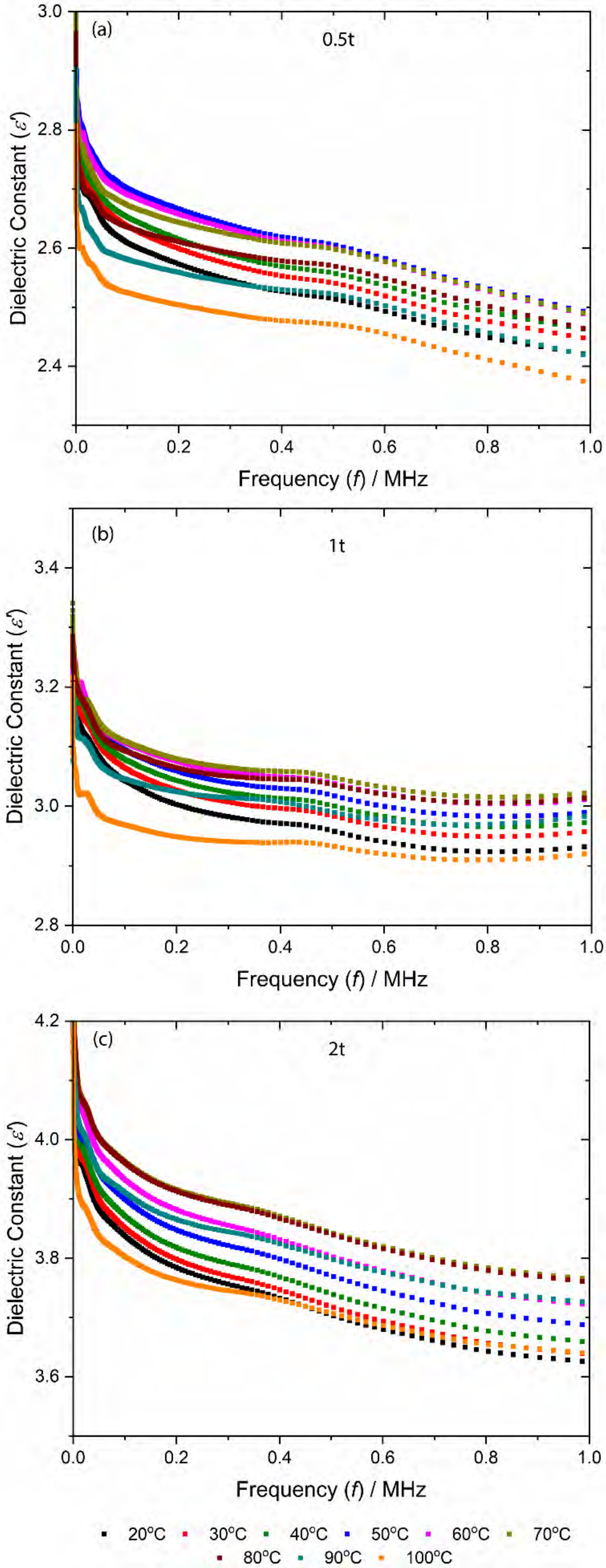

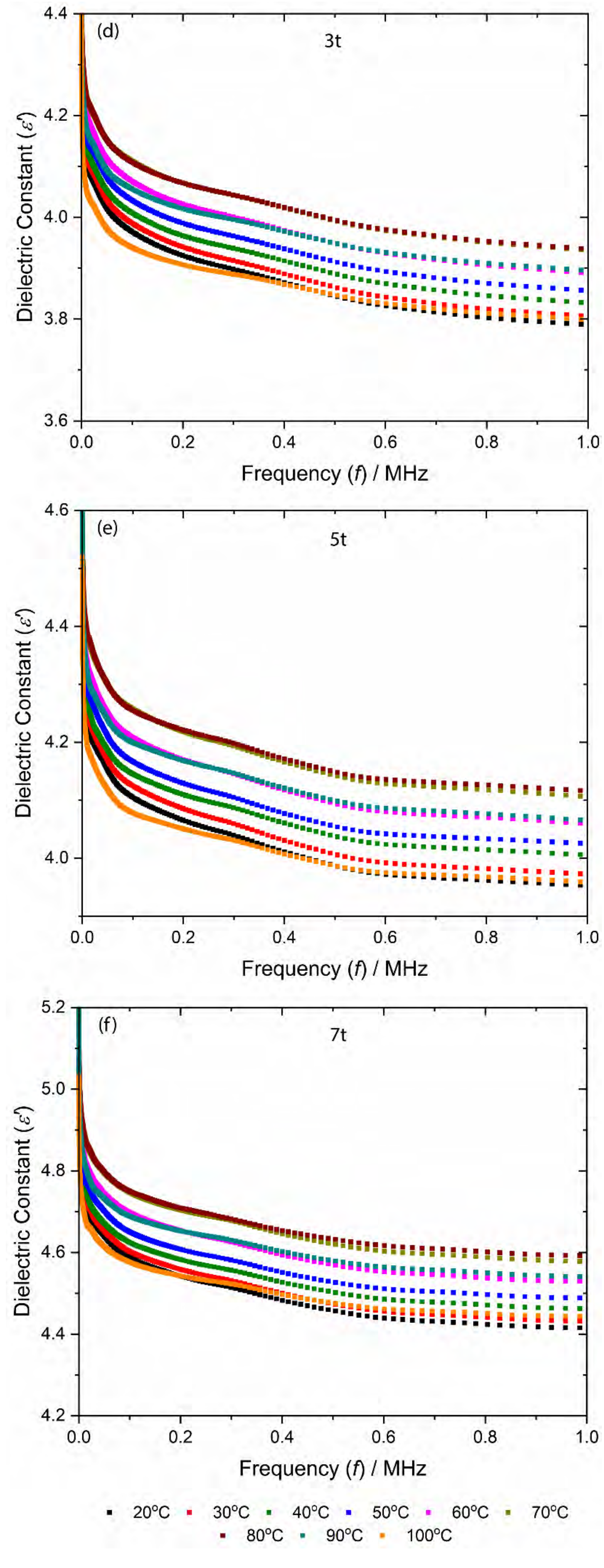
(d)
3t
Dielectric Constant (ε')
Frequency (f) / MHz
(e)
5t
Dielectric Constant (ε')
Frequency (f) / MHz
(f)
7t
Dielectric Constant (ε')
Frequency (f) / MHz
20°C 30°C 40°C 50°C 60°C 70°C
80°C 90°C 100°C

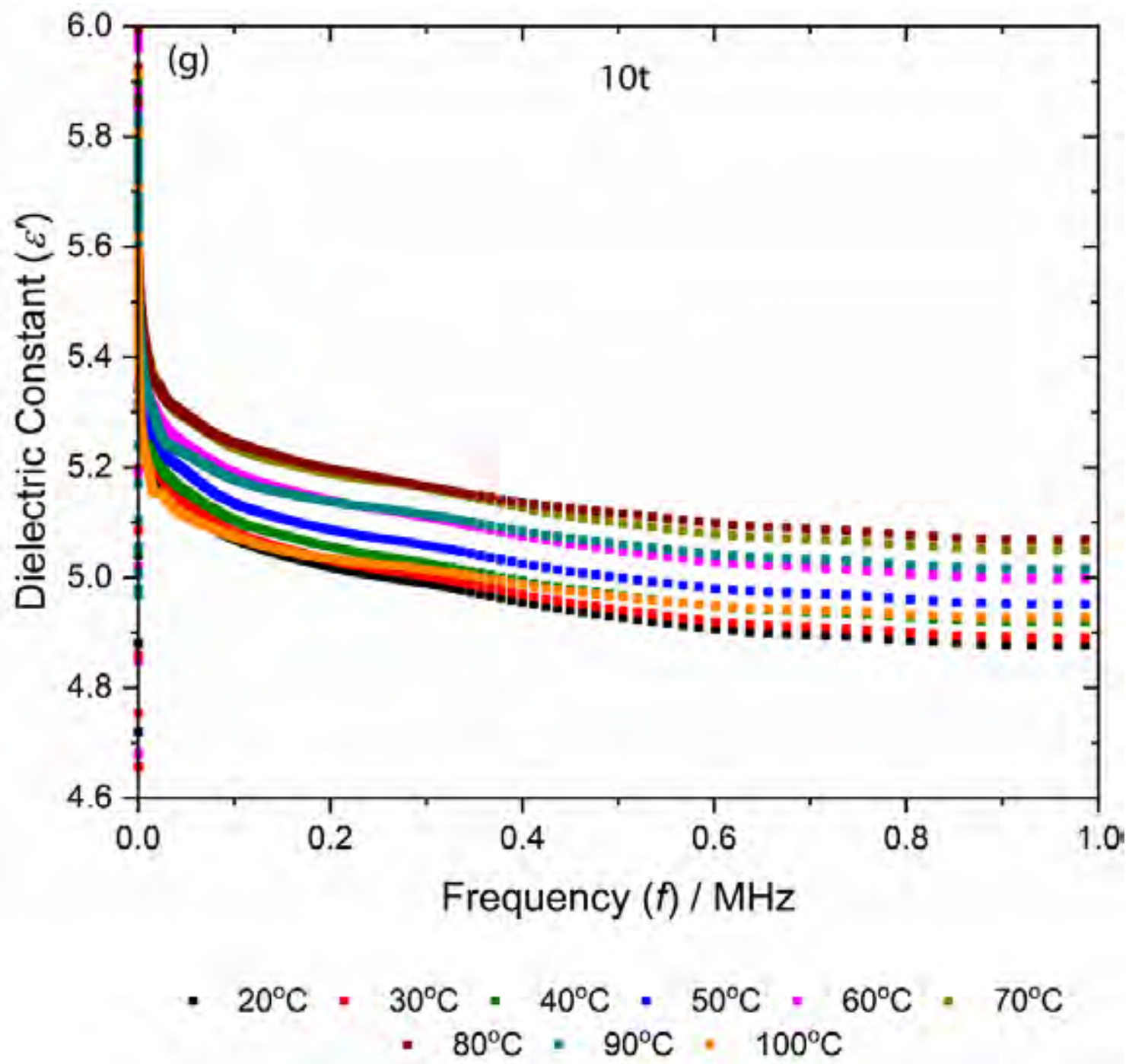

Figure S8: Temperature dependent real part of the dielectric constant as a function of frequency for HKUST-1 pellets prepared under a compression load of: (a) 0.5 ton, (b) 1 ton, (c) 2 ton, (d) 3 ton, (e) 5 ton, (f) 7 ton and (g) 10 ton, corresponding to the pressure of 36.96, 73.92, 147.84, 221.76, 369.60, 517.44 and 739.20 MPa.

## 8. Imaginary part of dielectric constant in MHz region - Individual pellet dielectrics

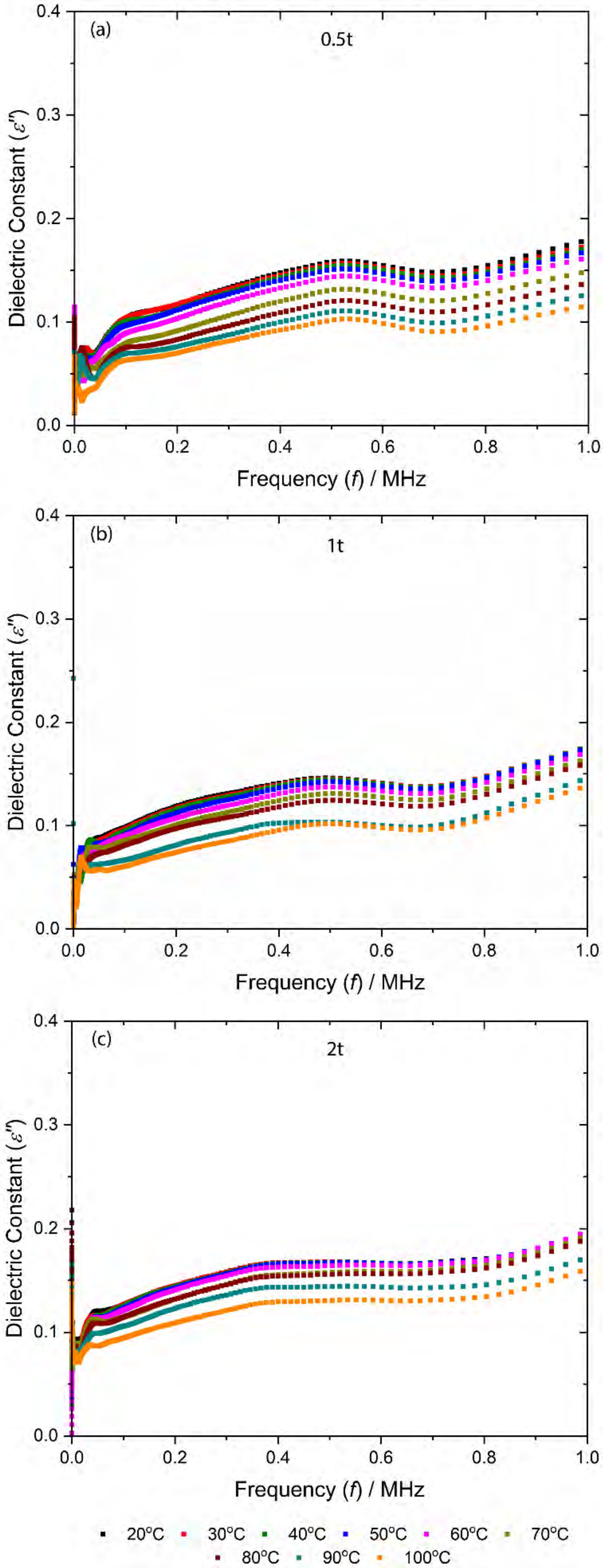

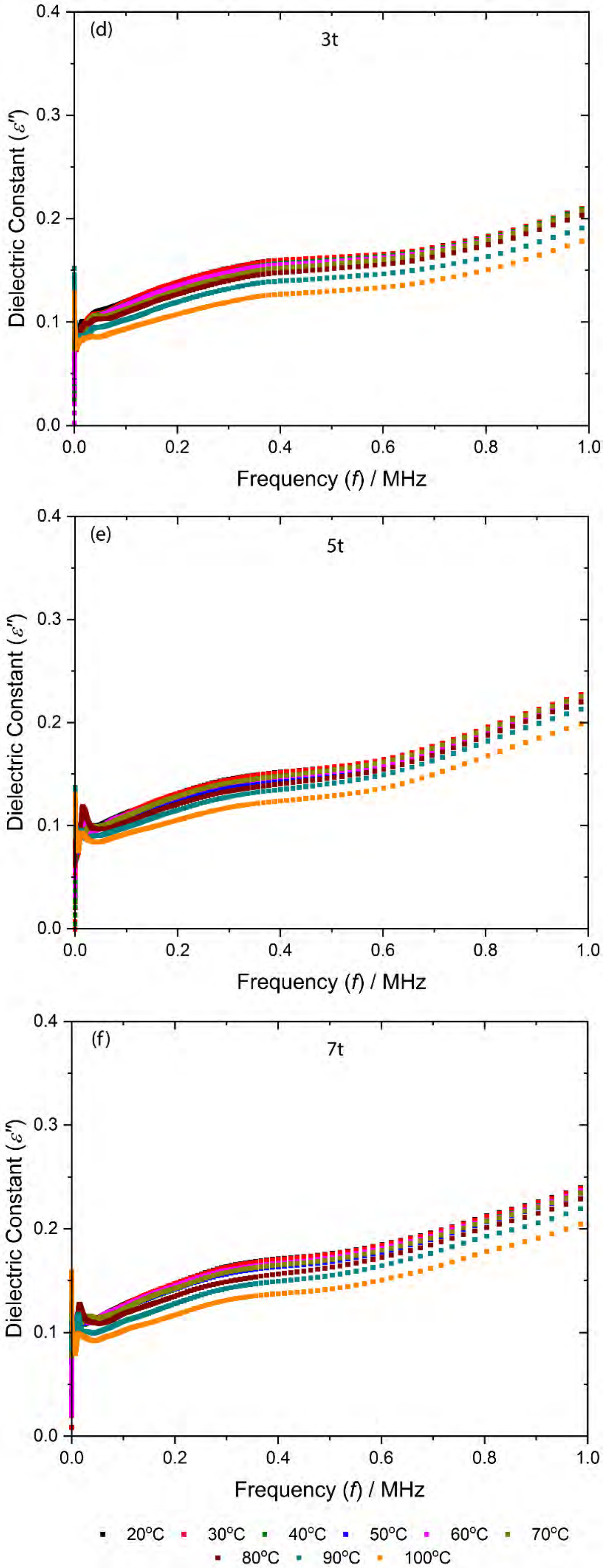
(d)
3t
Dielectric Constant (ε'')
Frequency (f) / MHz
(e)
5t
Dielectric Constant (ε'')
Frequency (f) / MHz
(f)
7t
Dielectric Constant (ε'')
Frequency (f) / MHz
20°C 30°C 40°C 50°C 60°C 70°C
80°C 90°C 100°C

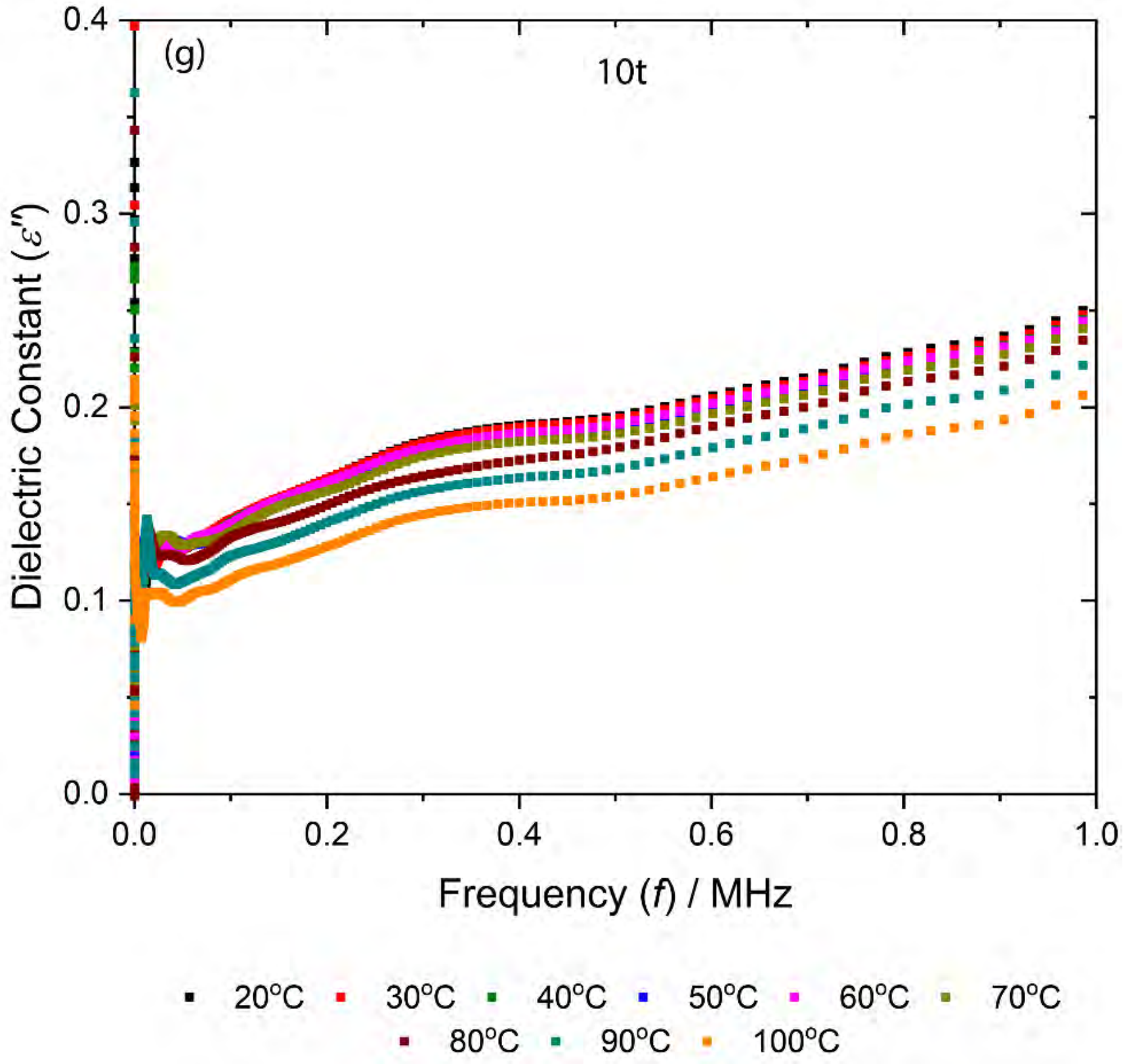

Figure S9: Temperature dependent imaginary part of the dielectric constant as a function of frequency for HKUST-1 pellets prepared under a compression load of: (a) 0.5 ton, (b) 1 ton, (c) 2 ton, (d) 3 ton, (e) 5 ton, (f) 7 ton and (g) 10 ton, corresponding to the pressure of 36.96, 73.92, 147.84, 221.76, 369.60, 517.44 and 739.20 MPa.

## 9. Loss tangent of dielectric constant in MHz region - Individual pellet dielectrics

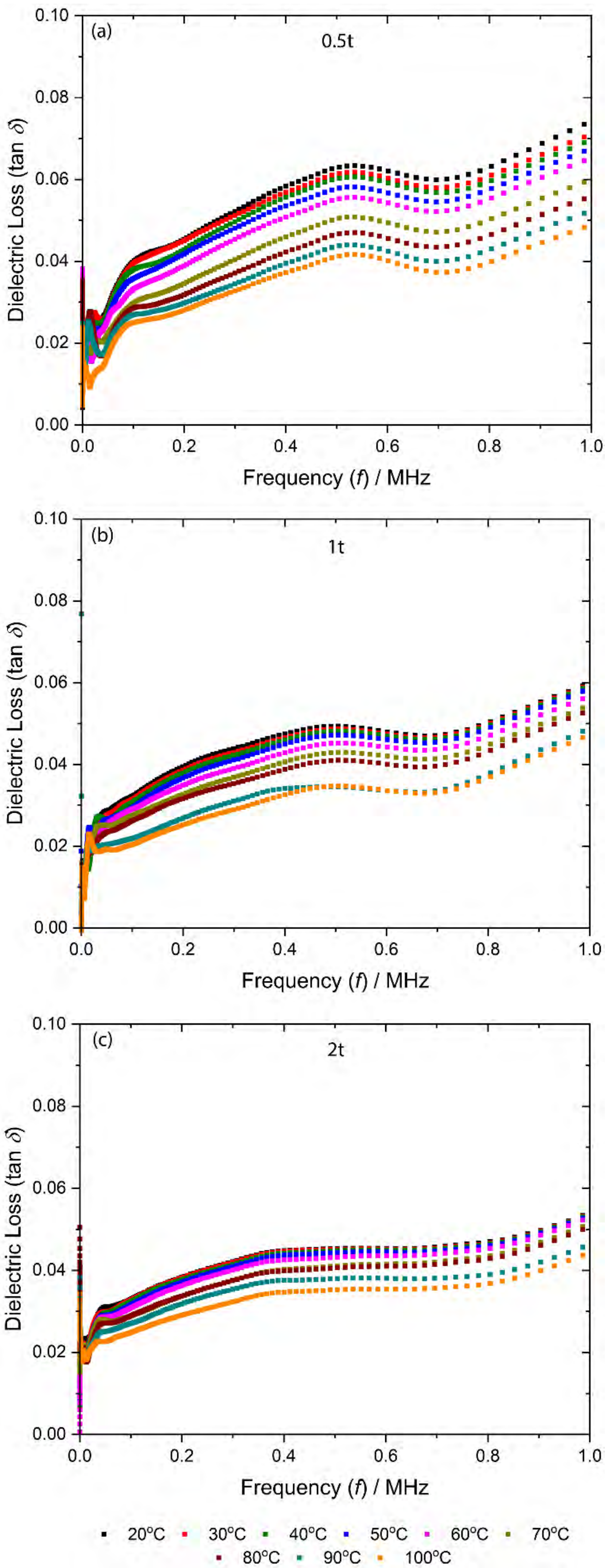

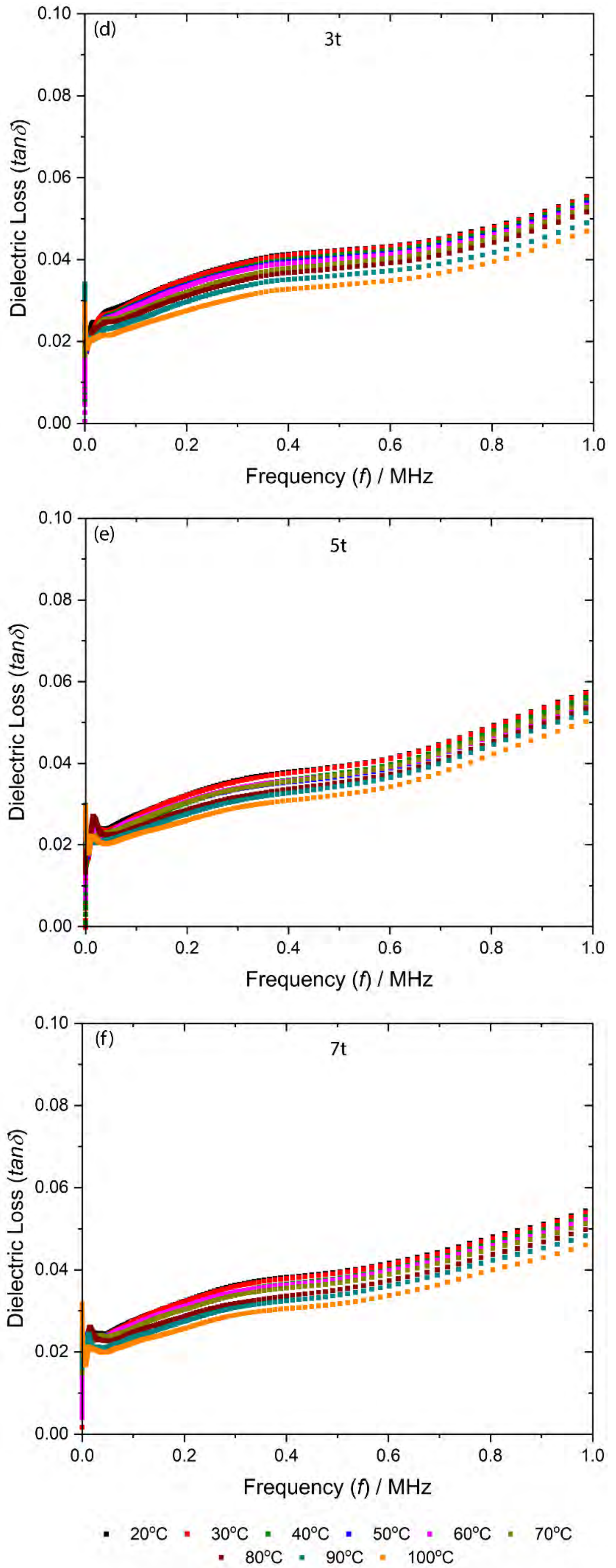
(d)
3t
Dielectric Loss (tanδ)
Frequency (f) / MHz
0.00
0.02
0.04
0.06
0.08
0.10
0.0
0.2
0.4
0.6
0.8
1.0
(e)
5t
Dielectric Loss (tanδ)
Frequency (f) / MHz
(f)
7t
Dielectric Loss (tanδ)
Frequency (f) / MHz
20°C
30°C
40°C
50°C
60°C
70°C
80°C
90°C
100°C

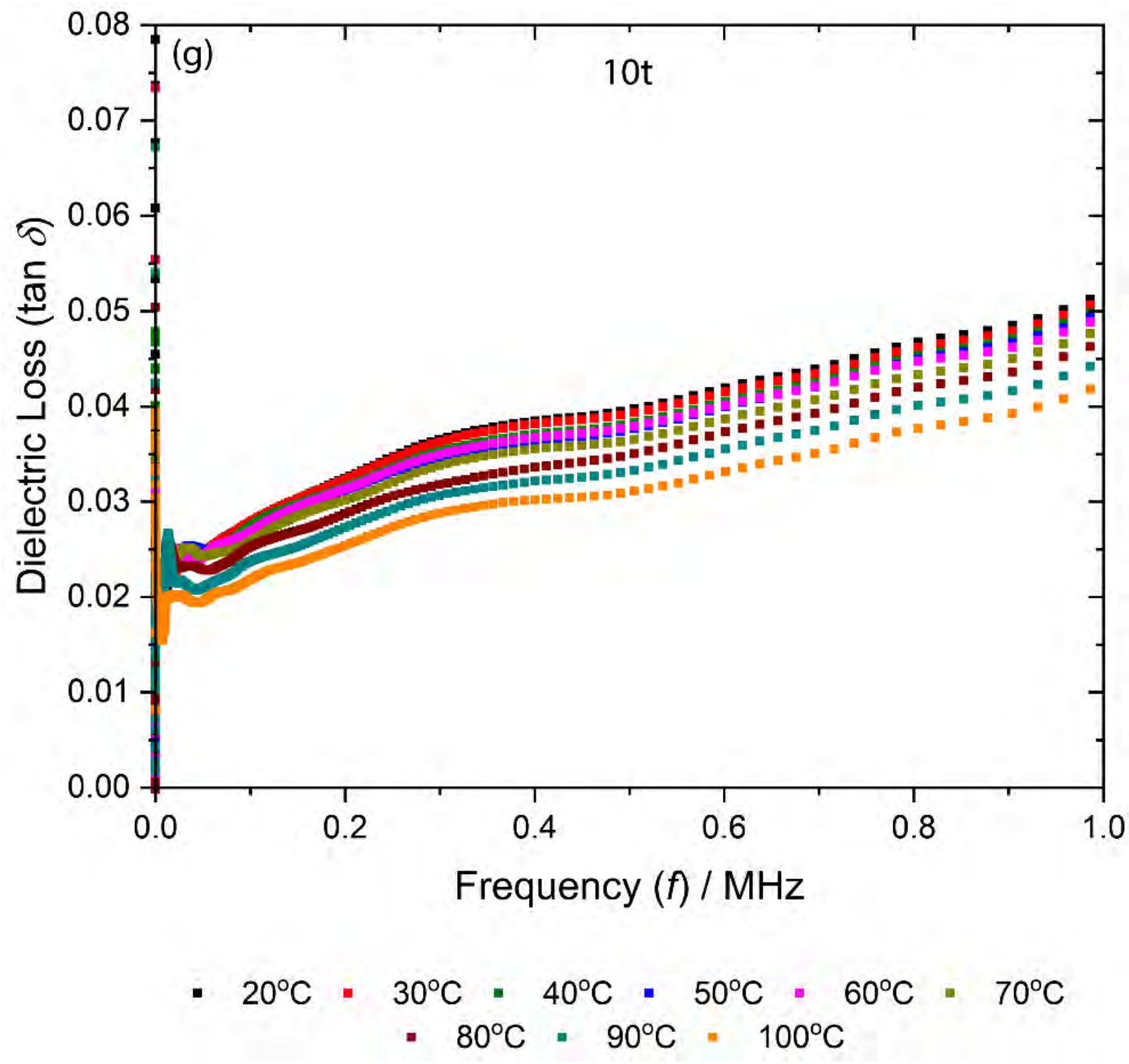

Figure S10: Temperature dependent dielectric loss as a function of frequency for HKUST-1 pellets prepared under a compression load of: (a) 0.5 ton, (b) 1 ton, (c) 2 ton, (d) 3 ton, (e) 5 ton, (f) 7 ton and (g) 10 ton, corresponding to the pressure of 36.96, 73.92, 147.84, 221.76, 369.60, 517.44 and 739.20 MPa.

# 10. Density functional theory (DFT) calculations of dielectric properties

Density functional theory (DFT) calculations were performed using the CRYSTAL17 code [5]. We used the B3LYP hybrid exchange-correlation functional [6] augmented with two- and three-body corrections for dispersive interactions [7] (*i.e.* B3LYP-D3(ABC)) in combination with a triple-zeta (TZP) quality basis set [8].

An ideal (defect free) HKUST-1 framework structure with ferromagnetic cubic symmetry (*Fm*-3*m*) was geometrically optimized in accordance with ref. [9].

The computational output contains the theoretical predictions of the reflectance spectrum and the complex dielectric function at 0 K for the optimized structure. The real and imaginary parts of the dielectric function were obtained through Kramers-Kronig relations [10] and plotted as the $\varepsilon'$ and $\varepsilon''$ values as a function of frequency.

Vibrational frequencies calculation and dielectric response were computed with additional CRYSTAL keywords listed below, to account for the contributions of nuclear motions to the static dielectric constants ($\omega = 0$).

```
FREQCALC
NUMDERIV
2
INTENS
INTCPHF
END
IRSPEC
REFRIND
DIELFUN
DAMPFAC
5.0
GAUSS
ENDIR
END
```

The pressure-dependent dielectric calculations at 190 MPa and 360 MPa were carried out by a two-step procedure. First, we run a series of constant-volume geometry optimization and then the data were fitted with a Birch-Murnagham (BM) equation of state. Second, the cell volume was fixed at the pressure corresponding to 190 MPa and 360 MPa as obtained from the BM fitting. Finally, vibrational frequencies and dielectric response were computed as described above. Note that the present calculations show that at 360 MPa the cubic structure is no longer stable and a phase transition occurs toward a lower symmetry tetragonal phase. Therefore, the reported results at 360 MPa correspond to the tetragonal unit cell.

## 11. Reflectivity spectra $R(\omega)$ in the far-IR and mid-IR regions

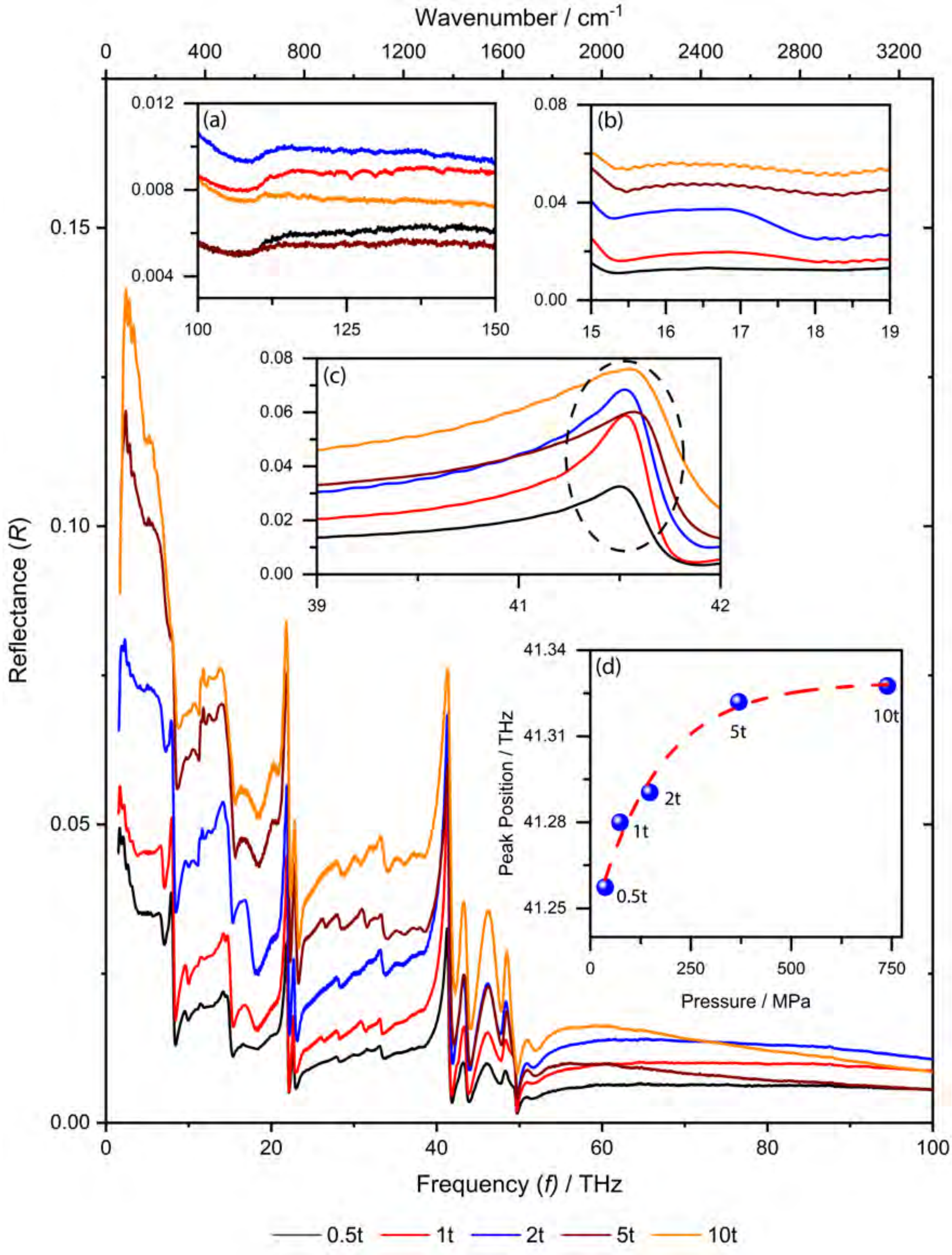

Figure S11: Reflectance spectra of HKUST-1 pellets. Inset: (a) reflectivity spectra from 100 - 150 THz, (b) both far-IR and mid-IR spectra are joined together at 600 $cm^{-1}$ (~18 THz), (c) and (d) shows the blue shift of peaks as a function of the pelleting pressure. The peak positions were identified using the Gaussian fit.

## 12. Real part of dielectric constant in THz region: experiments vs DFT

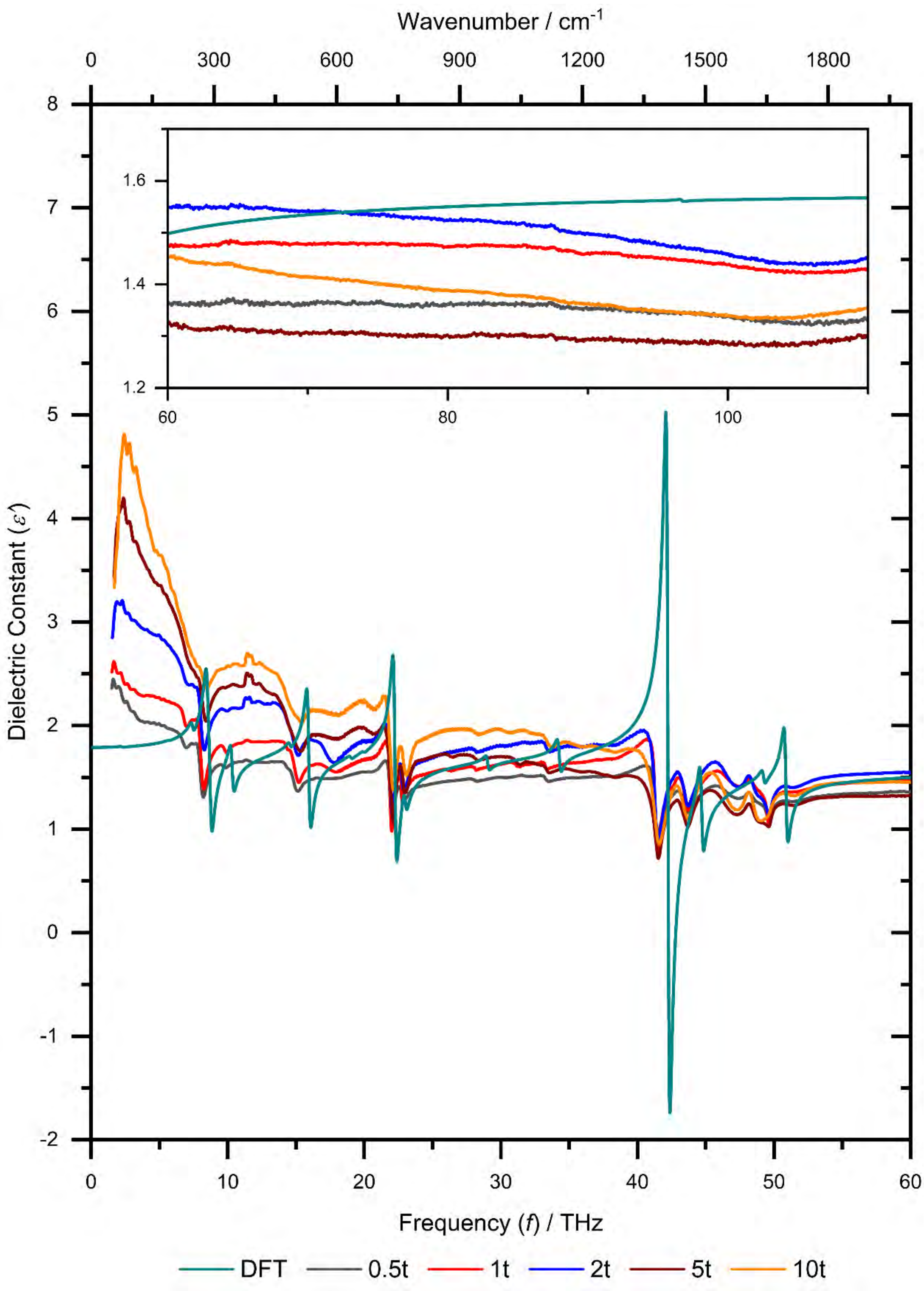

Figure S12: Overlapped experimental and simulated DFT spectra for the real part of dielectric constant as a function of frequency. The DFT calculation predicted all the oscillatory transitions in the HKUST-1 dielectric spectra. Inset shows the dielectric spectra at higher frequencies up to 150 THz.

## 13. Imaginary part of dielectric constant in THz region - experiments vs DFT

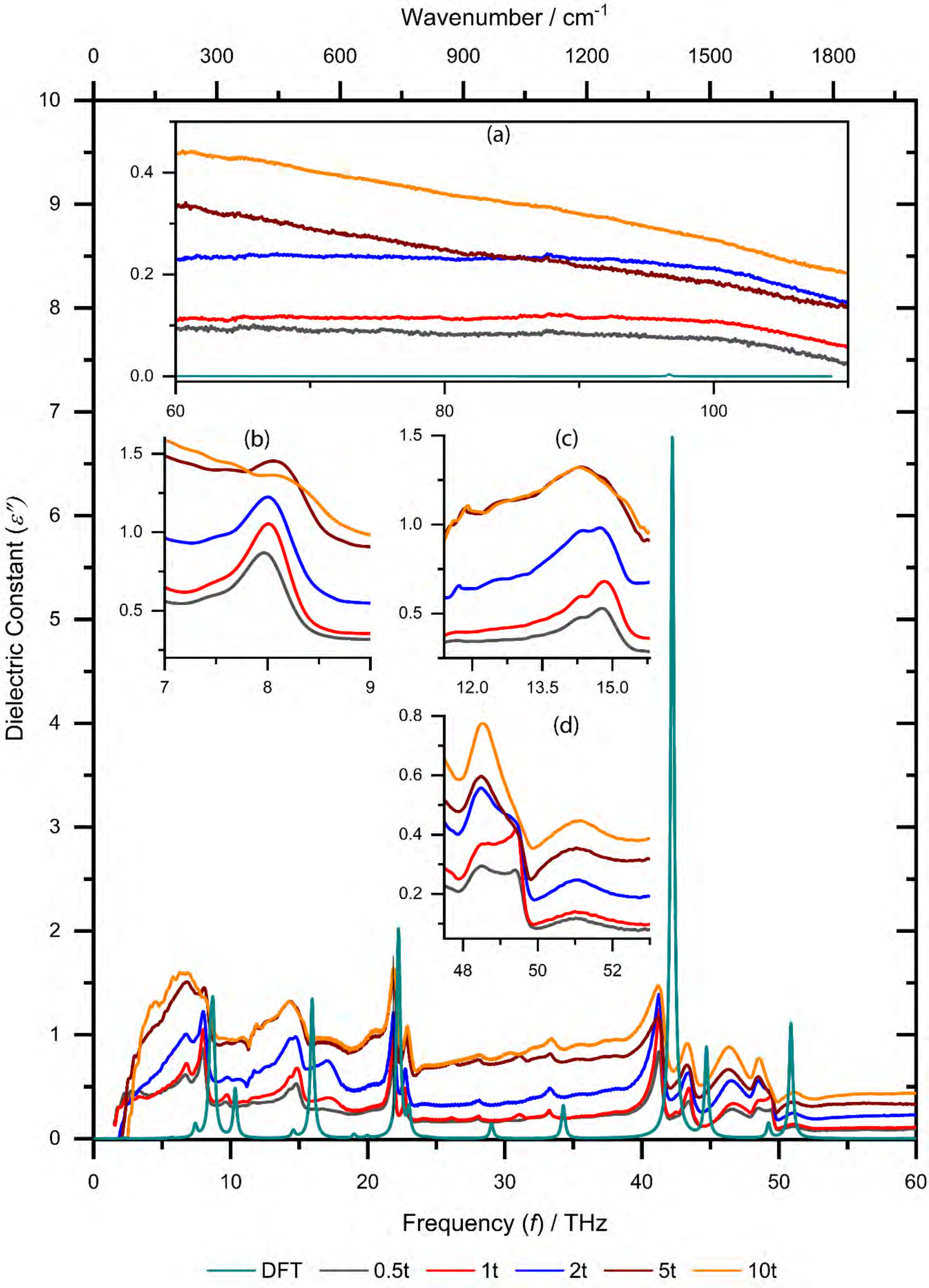

Figure S13: Overlapped experimental and simulated DFT spectra for the imaginary part of dielectric constant. Inset (a) shows the dielectric spectra at higher frequencies. The DFT

calculation predicted all the oscillatory transitions in the frequency-dependent dielectric spectra of HKUST-1. Interestingly, all the modes in the experimental data are slightly in the left side of the DFT simulated spectra. Inset (b) blue shift in the Cu-paddle wheel deformation mode (O-Cu-O bending and Cu-Cu buckling at 8 THz), with applying pressure, whereas inset (c) show the red shift in modes related to the linker deformation (out-of-plane aromatic ring deformation). Inset (d) demonstrate the blue shift in the mode associated with the Cu-O stretching.

## 14. Loss tangent (tan $\delta$) of HKUST-1 in MHz and THz regions

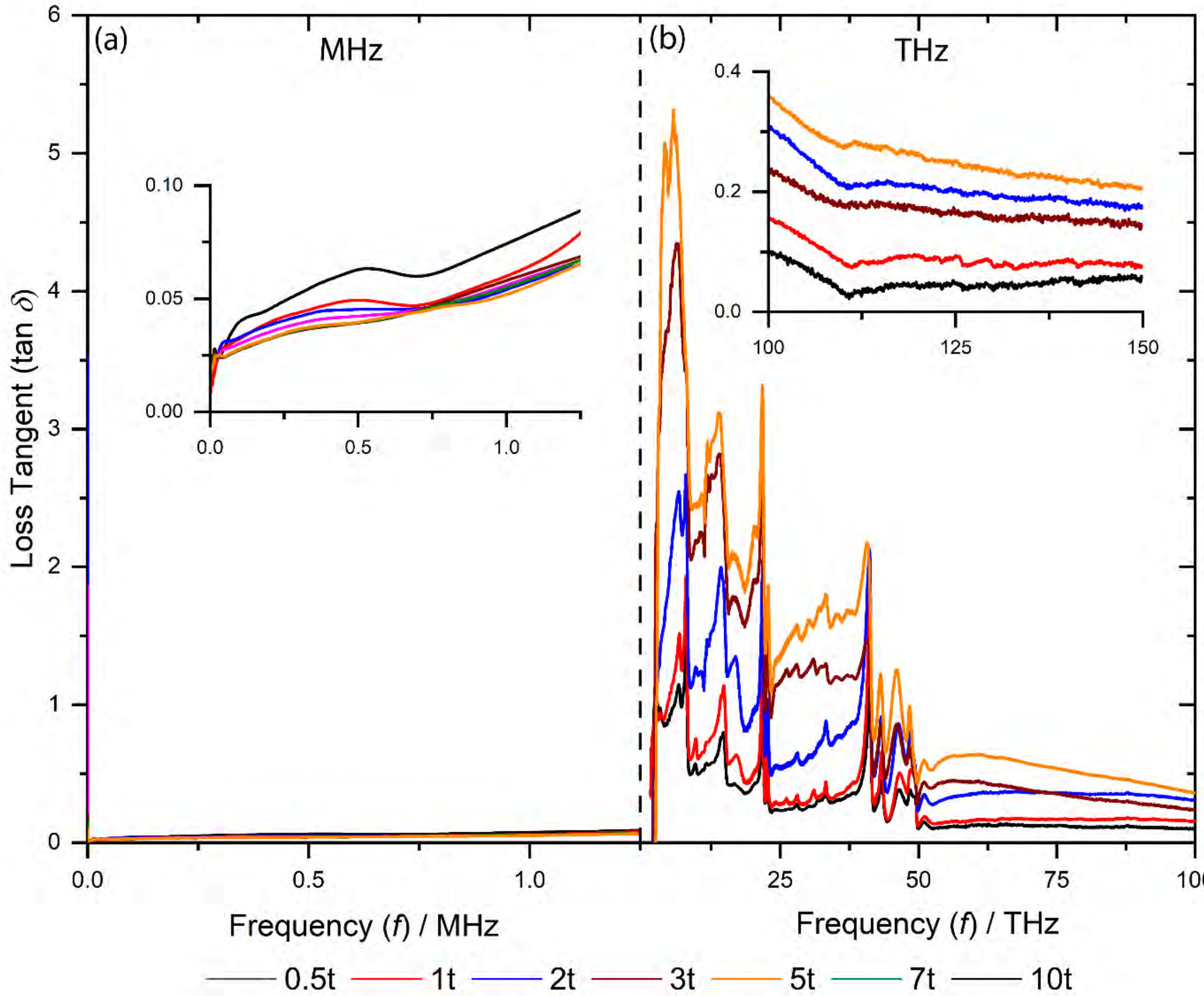

Figure S14: Dielectric loss tangent spectra of pellets in the MHz and THz region. The left inset shows the zoomed view of pellets in the MHz region, whereas right inset shows the extended dataset up to 150 THz.

# 15. Pressure-dependent DFT calculations

## 15.1 Components of the complex refractive index, *n* + *iκ*

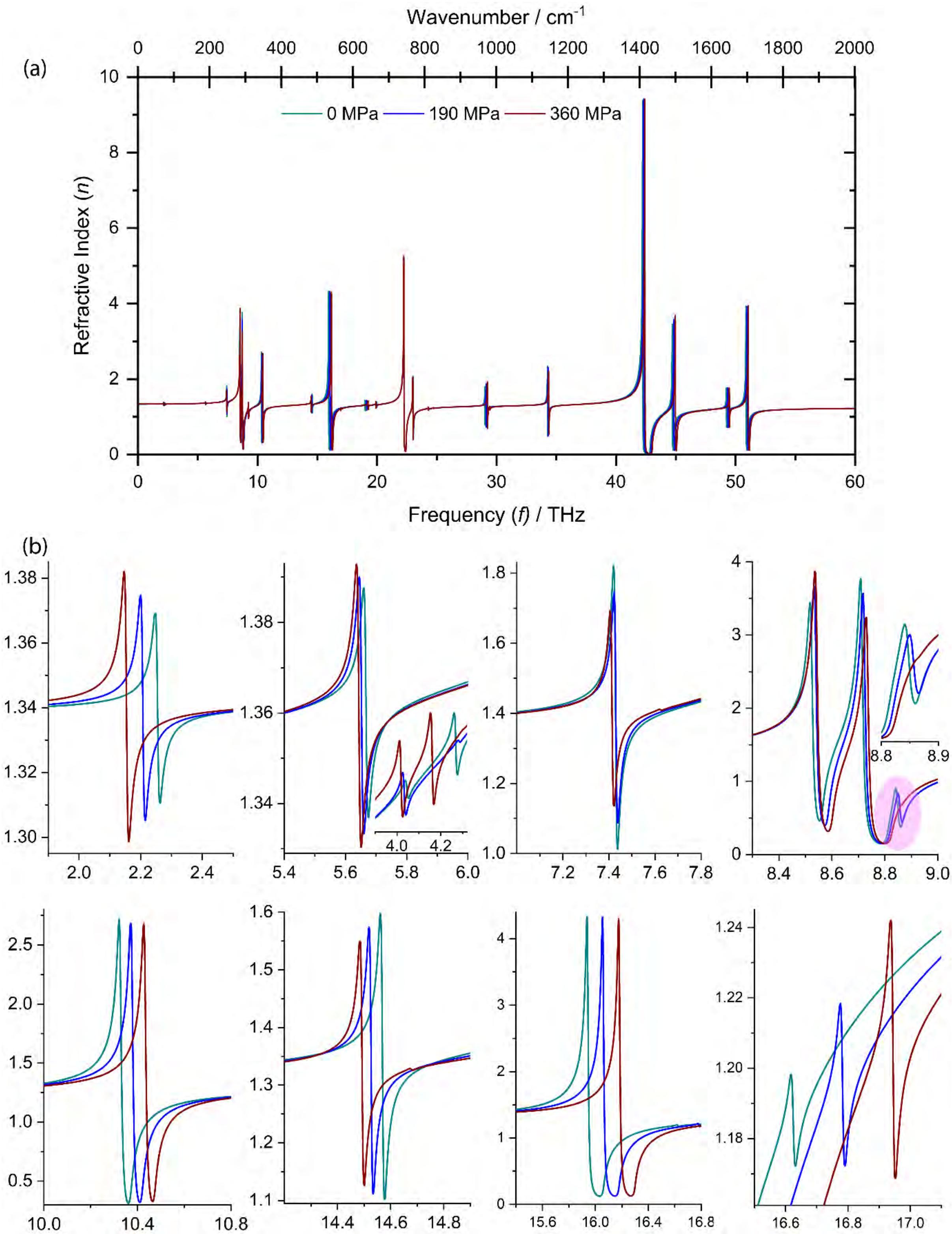

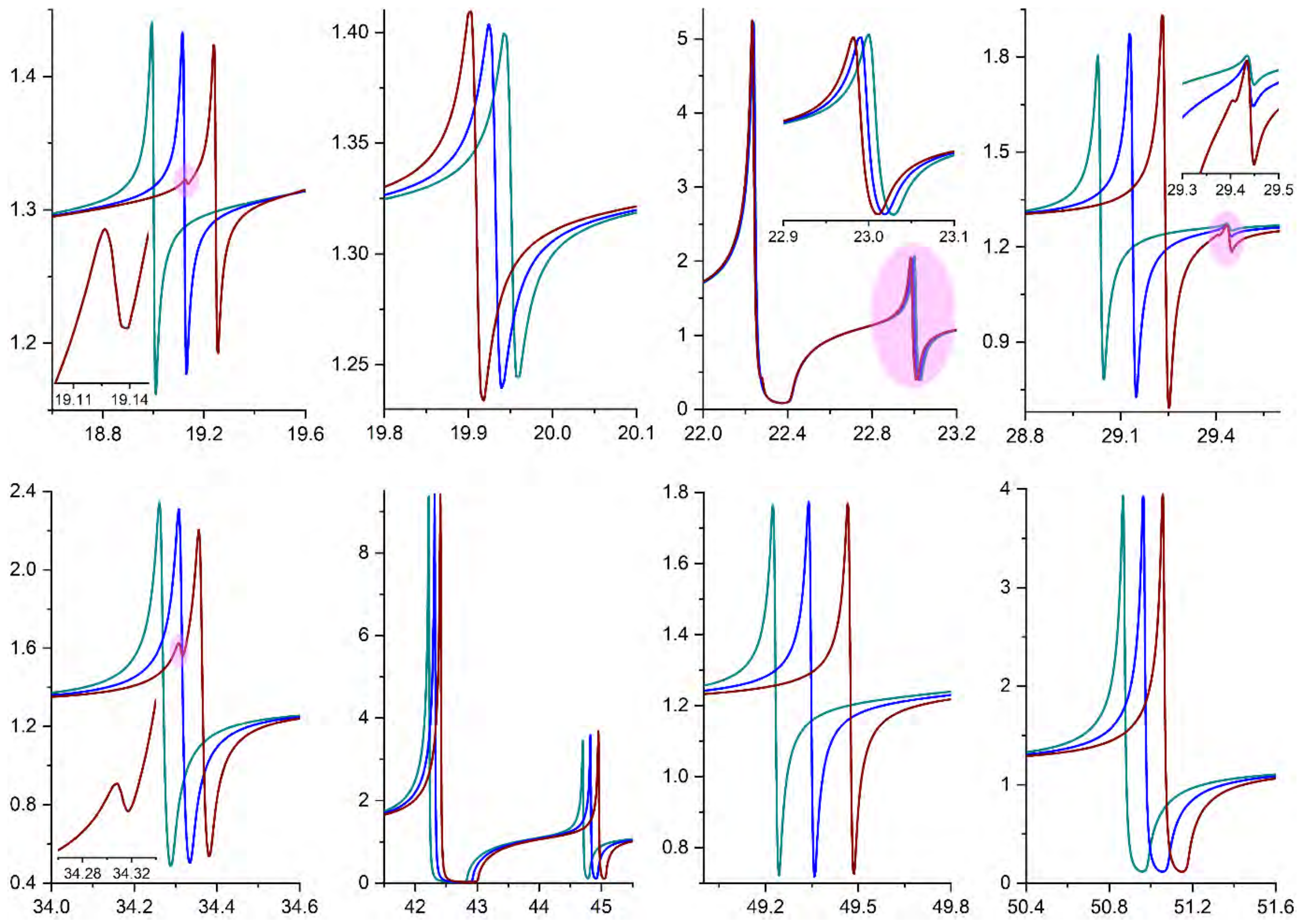

Figure S15: Pressure-dependent refractive index calculated by DFT. (a) Real part of the refractive index *n*. (b) Zoomed plots of the selected frequencies showing the red/blue shifts.

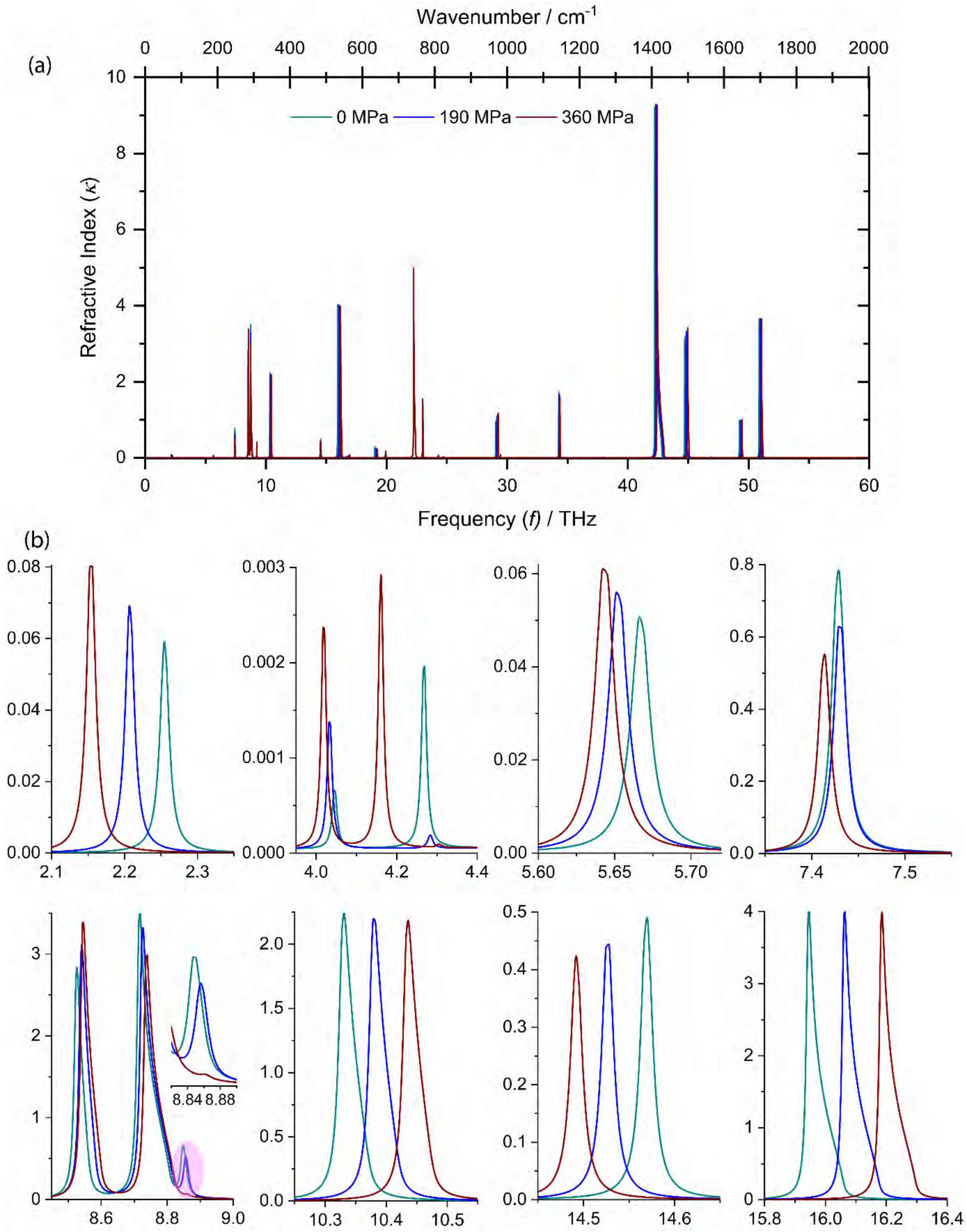
(a)
Wavenumber / cm-1
0 MPa
190 MPa
360 MPa
Refractive Index (κ)
Frequency (f) / THz
(b)

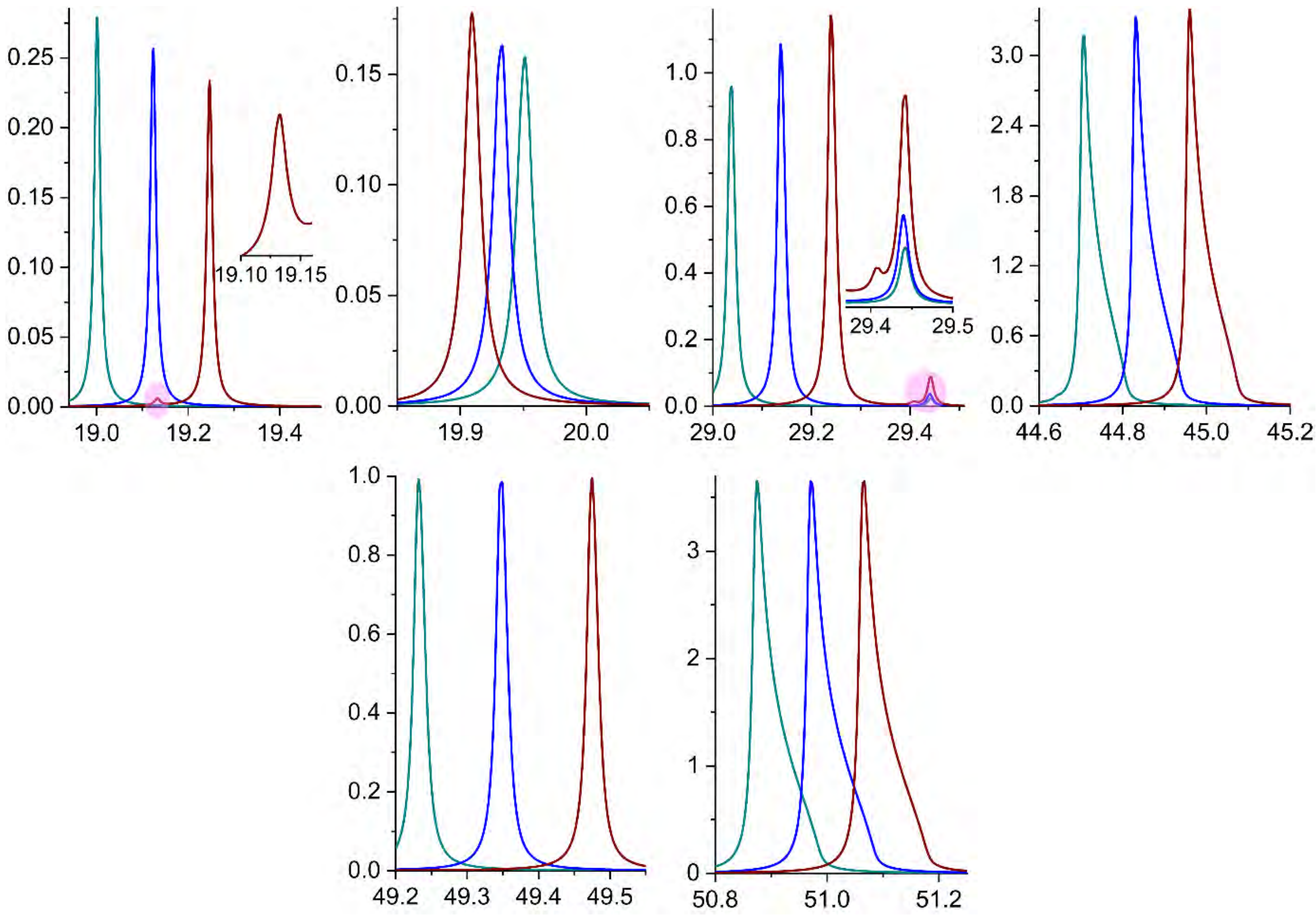

Figure S16: Pressure-dependent refractive index calculated by DFT. (a) Imaginary part of the refractive index, $\kappa$. (b) Zoomed plots of the selected frequencies showing the red/blue shifts.

## 15.2 Components of the complex dielectric property, $\varepsilon' + i\varepsilon''$

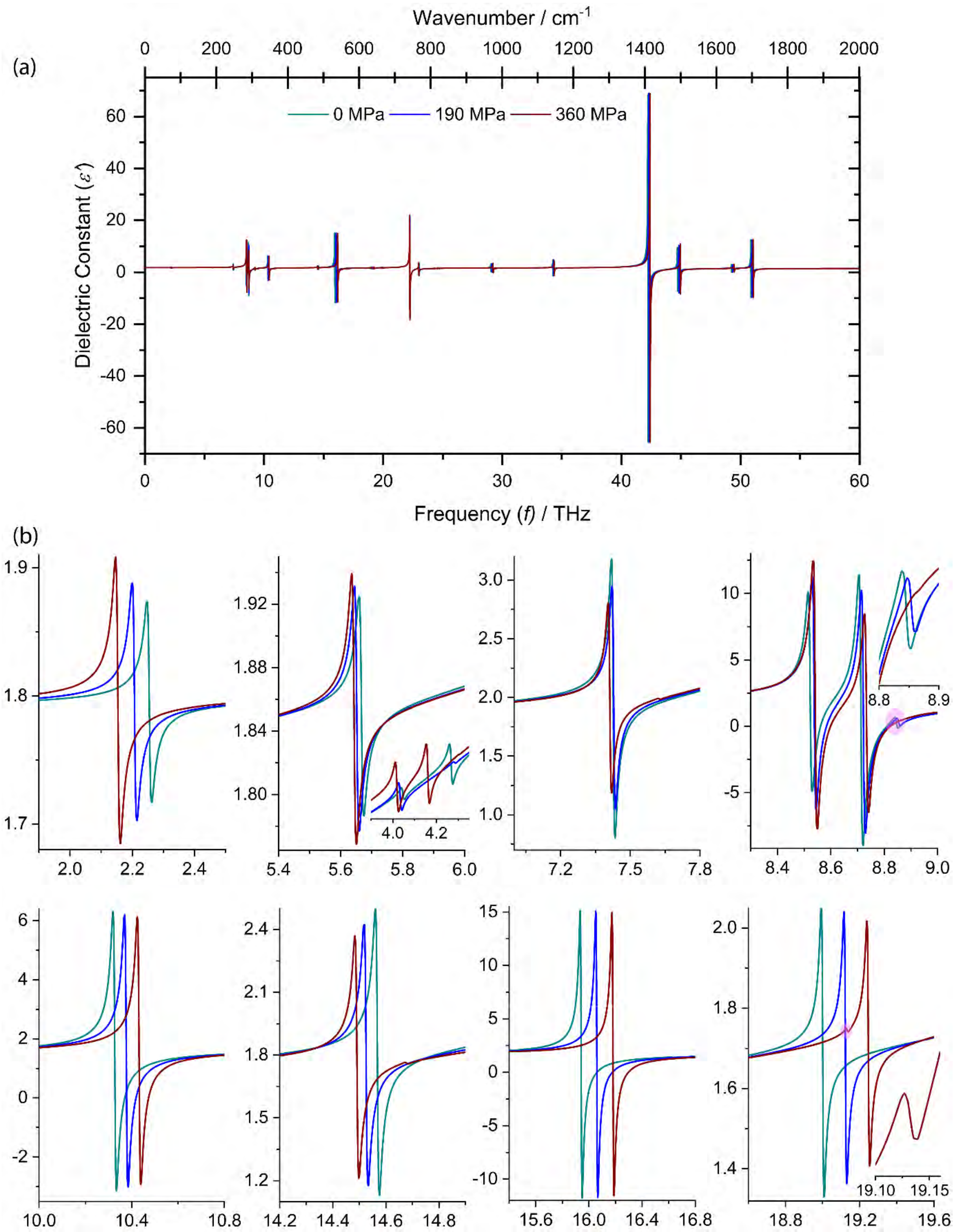

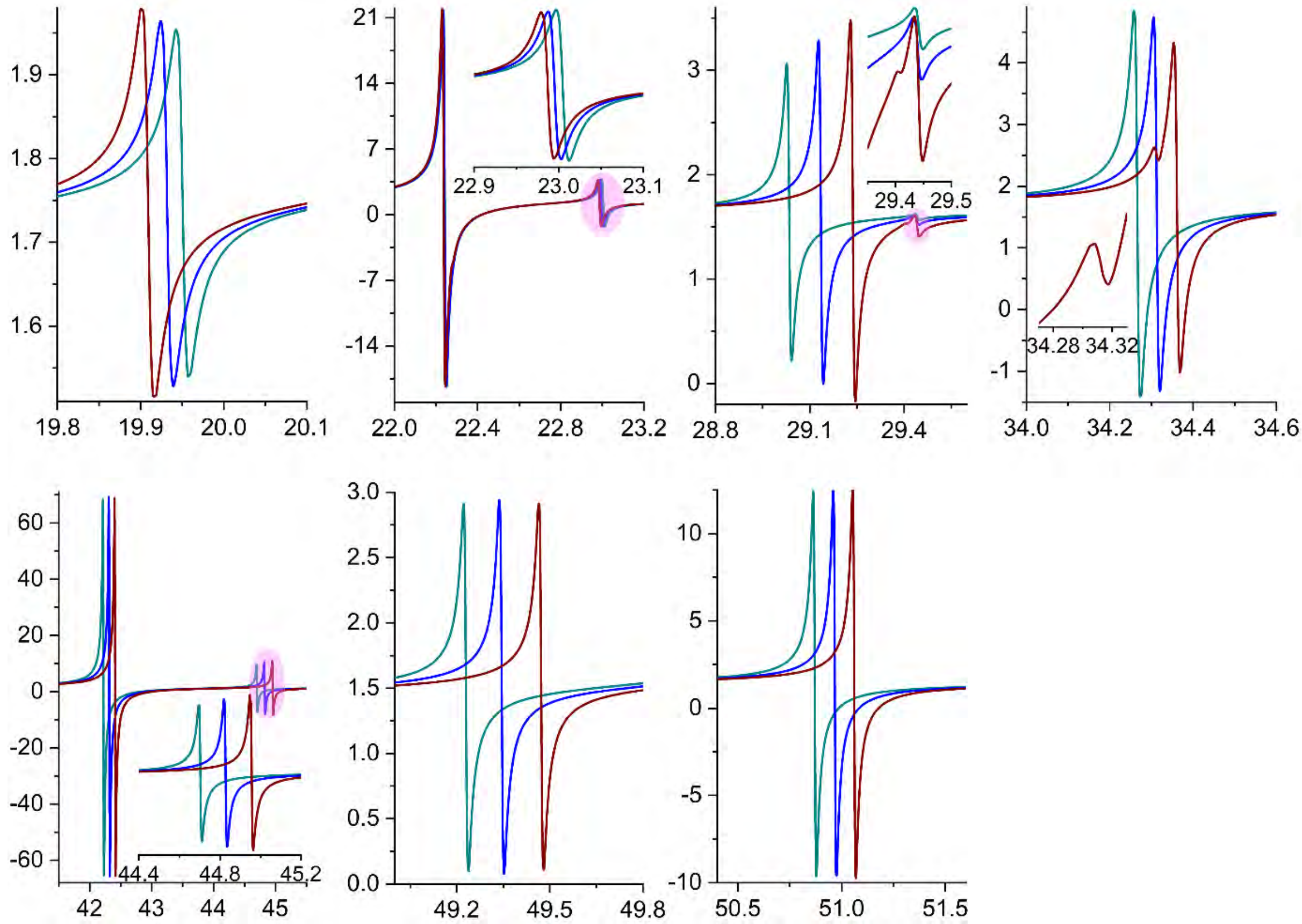

Figure S17: Pressure-dependent dielectric constant calculated by DFT. (a) Real part of the dielectric constant $\varepsilon'$. (b) Zoomed plots of the selected frequencies showing the red/blue shifts.

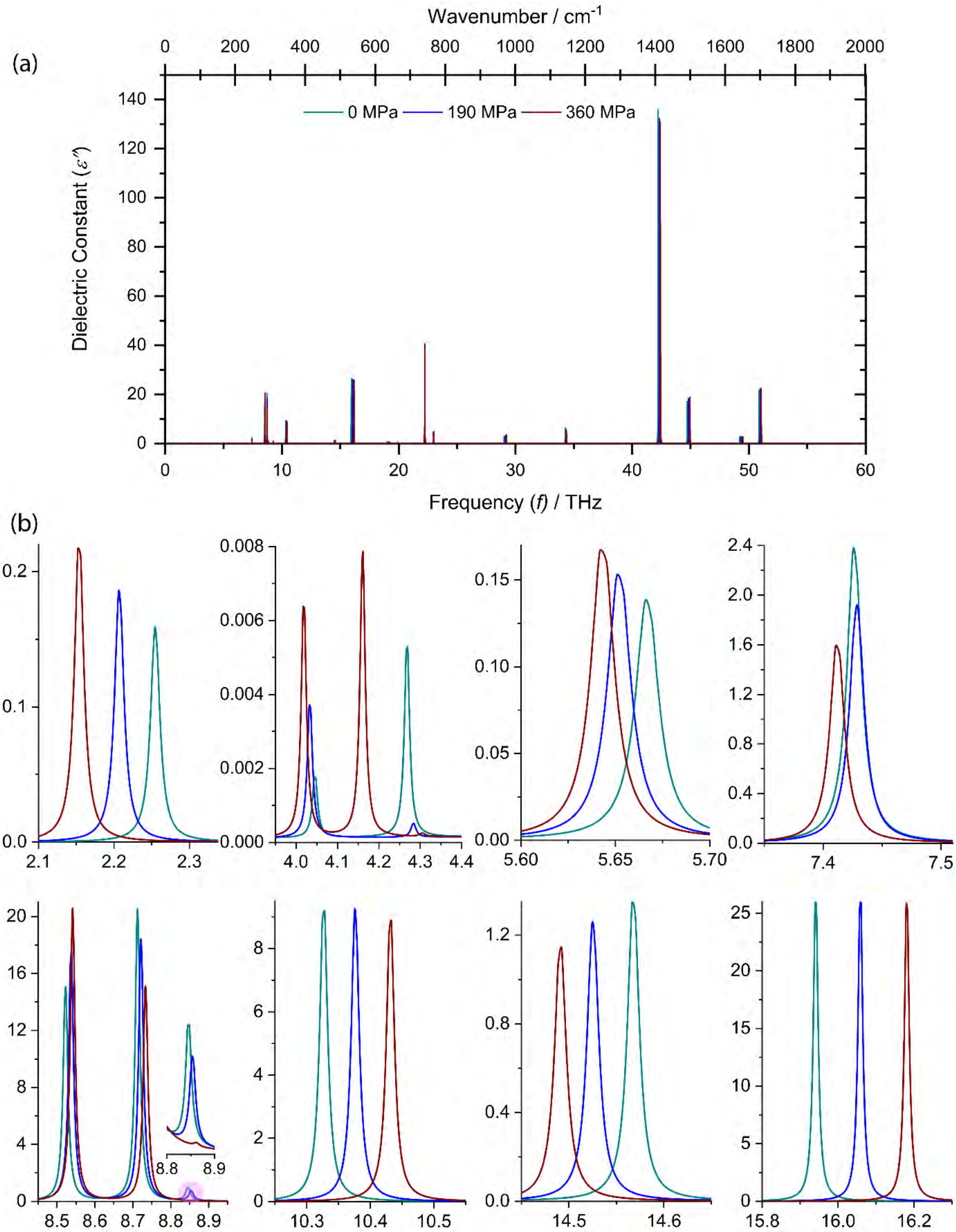
(a)
Wavenumber / cm-1
0 MPa
190 MPa
360 MPa
Dielectric Constant (ε′′)
Frequency (f) / THz
(b)

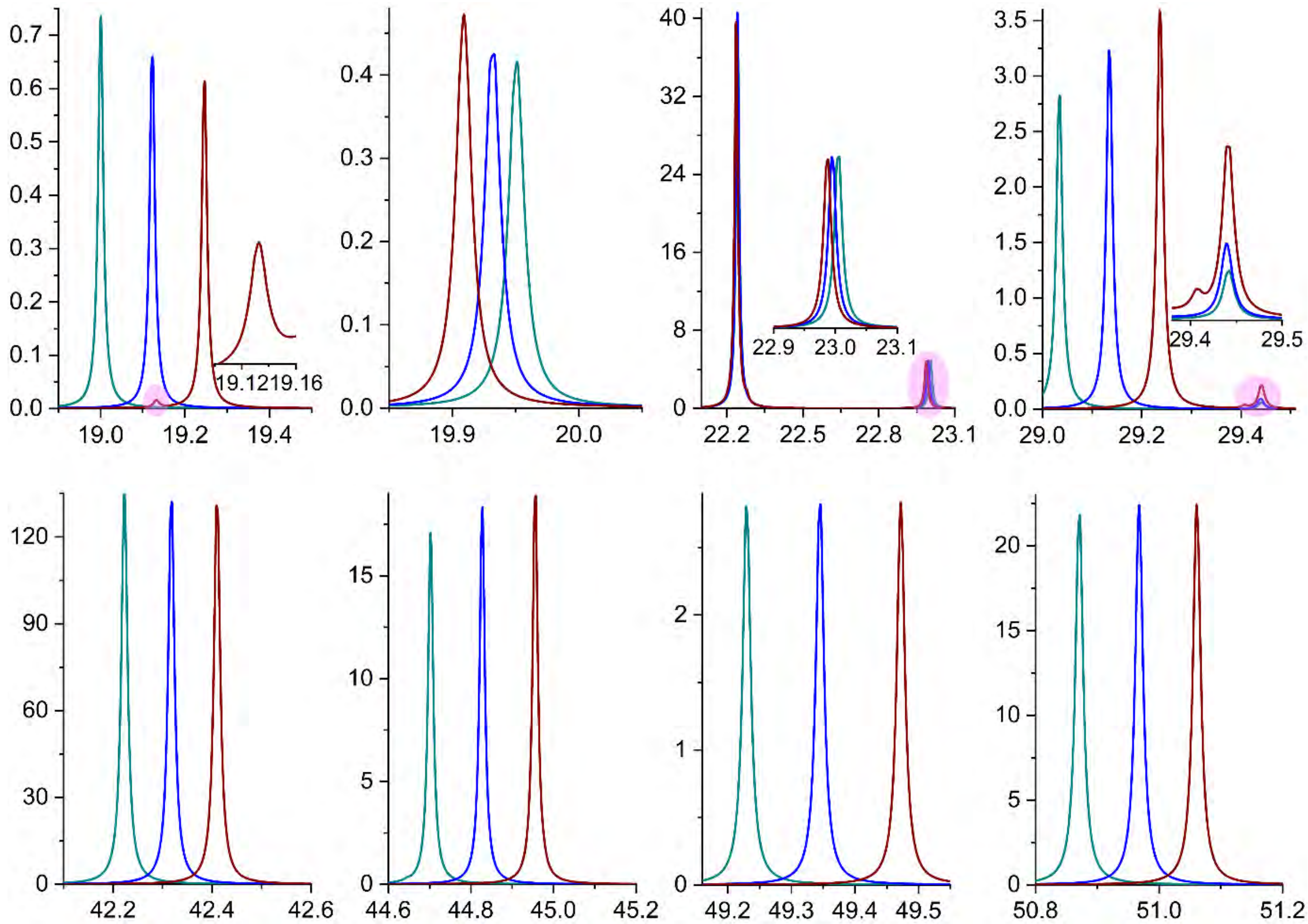

Figure S18: Pressure-dependent dielectric constant calculated by DFT. (a) Imaginary part of the dielectric constant $\varepsilon''$. (b) Zoomed plots of the selected frequencies showing the red/blue shifts.